\documentclass[a4paper,11pt]{article}
\pdfoutput=1 

\usepackage{jcappub} 
\usepackage[utf8x]{inputenc}                   
\usepackage[dvipsnames]{xcolor}
\usepackage{bbold}
\usepackage[T1]{fontenc} 
\usepackage{ulem}

\usepackage{tikz}
\usetikzlibrary{arrows,positioning} 
\tikzset{
    >=stealth',
    operator/.style={rectangle, rounded corners, fill=gray!10, minimum width=10.7em, minimum height=2em, align=center},
    freccia/.style={->, thick, shorten <=2pt, shorten >=2pt,}
}
\newcommand{\microboxwidth}{8.7cm}
\newcommand{\macroboxwidth}{3.7cm}
\newcommand{\boxheight}{8cm}
\pgfdeclareimage[width=\microboxwidth]{micro}{fig_microstates}
\pgfdeclareimage[width=\microboxwidth]{smearedSmall}{fig_smeared_microstates}
\pgfdeclareimage[width=\macroboxwidth]{macroSmall}{fig_macrostate}

\title{\boldmath The Halo Boltzmann Equation}

\author[a]{Matteo Biagetti,}
\author[a]{Vincent Desjacques,}
\author[a, b]{Alex Kehagias,}
\author[a]{Davide Racco,}
\author[a]{Antonio Riotto}

\affiliation[a]{
D\'epartement de Physique Th\'eorique and Center for Astroparticle Physics, \\ 
Universit\'e de Gen\`eve, 24 quai Ansermet, CH-1211 Gen\`eve 4, Switzerland}
\affiliation[b]{Physics Division, National Technical University of Athens,15780 Zografou Campus, Athens, Greece}

\emailAdd{matteo.biagetti@unige.ch}
\emailAdd{vincent.desjacques@unige.ch}
\emailAdd{kehagias@central.ntua.gr}
\emailAdd{davide.racco@unige.ch}
\emailAdd{antonio.riotto@unige.ch}

\usepackage{xcolor}
\definecolor{MatteoColour}{rgb}{0.,0.7,0.3}
\definecolor{vincentColour}{rgb}{0.,0.05,0.9}
\definecolor{AlexColour}{rgb}{0.7,0.05,0.7}
\definecolor{DavideColour}{rgb}{0.9,0.4,0.1}
\definecolor{ToniColour}{rgb}{0.1,0.3,0.}

\definecolor{ZeroColour}{rgb}{0.25,0.25,0.8}
\newcommand{\zero}[1]{\textcolor{ZeroColour}{#1}}

\def\d{{\rm  d}} 
\def\p{\partial}
\def\vk{{\vec k}}
\def\vq{{\vec q}}
\def\vp{{\vec v}}
\def\vr{{\vec r}}
\def\vl{{\vec \ell}}
\def\vv{{\vec v}}
\def\vx{{\vec x}}

\def\vpsi{\vec{\psi}}
\def\fpk{f_\text{pk}}

\def\vnabphi{\vec \nabla\Phi}
\def\vnabdelta{\vec \nabla \delta}
\def\Otwo{\delta\vnabphi}
\def\Othree{(\vnabphi \hspace{-2pt} \cdot \hspace{-2pt} \vec\nabla)\vnabphi}
\newcommand{\corr}[2]{\langle \mathcal O_{#1} \mathcal O_{#2}\rangle}
\newcommand{\corrpk}[2]{\langle \mathcal O_{#1} \mathcal O_{#2}\rangle_\text{pk}}
\newcommand{\ren}[1]{\left[ #1 \right]}
\def\sss{\sigma^4_0-\sigma^2_1\sigma^2_{-1}}

\def\rpt{(\vec{r},\vec{v},t)}

\newcommand{\vrp}{\vec{r}^{\, \prime}}
\newcommand{\vvp}{\vec{v}^{\, \prime}}
\def\rt{(\vr,t)}
\def\rvt{(\vr,\vv,t)}
\def\rpvt{(\vec{r},\vec{v}_2,t)}
\newcommand{\dvfrvt}{\partial_\vv f \rvt}
\newcommand{\dvf}{\partial_\vv f }
\newcommand{\ww}{_R}

\def\be   {\begin{equation}}   \def\ee   {\end{equation}}
\newcommand{\muABB}{\mu_{\delta'\delta\delta}}
\newcommand{\muABC}{\vec\mu_{\delta'\delta \vec\psi}}
\newcommand{\muABD}{\vec\mu_{\delta'\delta \vec\nabla\delta}}
\newcommand{\muACC}{\mu_{\delta'\vec\psi \vec\psi}}
\newcommand{\muACD}{\mu_{\delta'\vec\psi \vec\nabla\delta}}
\newcommand{\muADD}{\mu_{\delta'\vec\nabla\delta \vec\nabla\delta}}
\newcommand{\muBBB}{\mu_{\delta\delta\delta}}
\newcommand{\muBCC}{\mu_{\delta\vec\psi \vec\psi}}
\newcommand{\muBCD}{\mu_{\delta\vec\psi \vec\nabla\delta}}
\newcommand{\muBDD}{\mu_{\delta\vec\nabla\delta \vec\nabla\delta}}

\abstract{Dark matter halos are the building blocks of the universe as they host galaxies 
and clusters. The knowledge of the clustering properties of  halos is therefore essential 
for  the understanding of the  galaxy statistical properties.  
We derive an effective halo  Boltzmann equation which can 
be used to describe the halo clustering statistics. In particular, we show how the halo Boltzmann equation encodes a statistically biased gravitational force which generates
a bias in the peculiar  velocities of virialized halos with respect to the underlying dark matter, as recently observed in N-body simulations.}

\begin{document} 
\maketitle
\flushbottom

\section{Introduction}
\label{sec:intro}

\hskip 0.8cm Over the last decade, we have accumulated a good deal of observational evidence that 
the large scale structure (LSS)   originated from seed fluctuations in the very early universe, 
which grew  via gravitational instability at later times. 
In this picture, galaxies form within Dark Matter (DM) halos \cite{whiterees}, and the  goal of 
the theory of bias \cite{kaiser84} is to relate observable properties of tracers such as the density 
contrast of galaxies to the underlying matter distribution and, ultimately, to the initial conditions.

DM  halos are, to a large extent, the building blocks of what we observe. Knowing  how DM  halos 
form and evolve under the action of gravity is crucial to understand  the environment that 
harbors galaxies and clusters. Understanding the clustering properties of halos
represents, therefore, a  key ingredient towards an accurate description of galaxy clustering 
statistics. 

The aim of this paper is to write down an Effective Boltzmann Equation (EBE) for the single phase 
space density of halo centers, which can be used to derive {\it ensemble average} halo clustering
statistics. Therefore, this must be an effective description, in which the halo distribution 
should be thought of as being the mean-field distribution associated with a given realization of the DM  
distribution. 
This is the correct interpretation of the halo overabundance defined through a perturbative bias expansion 
so long as the surveyed volume remains finite, as is the case of real or simulated data
\cite{dcss,Lazeyras:2015giz}.

Since the halo Boltzmann  equation will describe the dynamics of the halo mean-field, one should expect 
to recover exactly the Boltzmann equation for DM only if the halo centers are statistically unbiased 
relative to the coarse-grained DM density field. 
This is assumed to be true in the coupled-fluids approximation for the coevolution of DM and halos, 
which is largely used to compute the time evolution of bias by following the evolution over cosmic 
time and in Eulerian space of the halo progenitors - the so-called proto-halos - until their 
virialization \cite{fry,teg,elia,chan0}. 
In this framework, the momentum equations for halos and DM do not differ at all. However, this 
approximation is not valid in general, as one should expect that the force acting on large scale 
structure tracers is indeed biased relative to the force acting on a random field point. 
In other words, the sampling bias induced by the tracers propagates into their dynamical evolution.

Therefore, the EBE for halos will be different from that of DM.  
This point should not be underestimated: the DM Boltzmann equation is the starting point of
any analytical approach to the description of the  LSS (see e.g. \cite{davispeebles}).
Here we advocate that, since the ultimate goal is the understanding of the statistical 
properties of what we observe (that is galaxies), one ought  to adopt the EBE discussed below as the 
starting point for halo centers.

This is not the only reason for pursuing this project, though. 
LSS  probes such as peculiar velocities or redshift space distortions can be used to test the law of 
gravity through a measurement of peculiar velocities, i.e. deviations from pure Hubble flow 
\cite{vjd,kaiser87}. 
However, peculiar velocities are effectively measured at the position of halo centers. Therefore, even 
if the latter locally flow with the DM, velocity statistics may be biased if the halo peculiar velocities 
do not provide a fair sample of the matter flows 
(for recent discussions  see Refs. \cite{jennings,zzj1,jennings2,yzzj,jennings3,chiang}).
Thus far, measurements of the halo velocity power spectrum appear consistent with little or no 
statistical velocity bias \cite{pw,zzj1}. However, \cite{v1,jennings} reached somewhat different 
conclusions. Furthermore, the effect must vanish in the limit $k\to 0$ so that, at low redshift, 
scale-dependent nonlinearities complicate the interpretation of the results.

Peak theory predicts the existence of such a statistical effect which, in 
linear theory, manifests itself as a reduced halo velocity dispersion 
\cite{bbks} together with a $k$-dependent velocity bias in the usual Kaiser formula
\cite{v0} and in the time evolution of the linear peak bias \cite{dcss}.
The authors of Ref. \cite{v1} have recently detected through N-body simulations a 
large scale halo velocity bias $b_v(k)$ 

\be
\label{biasv}
b_v(k)=\bigl(1-R_v^2k^2\bigr),
\ee
which appears to remain constant throughout time until virialization. Here, 
$R_v$ is the typical scale of the halo velocity bias

\be
R_v^2=\frac{\sigma_0^2}{\sigma_1^2},\,\,\,\,\sigma_j^2=\int\frac{\d^3 k}{(2\pi)^3}\, k^{2j} 
P(k) \tilde{W}^2(kR),
\ee 
$P(k)$ is the DM power spectrum and $\tilde{W}(x)$  is a spherically symmetric, time-independent 
kernel with filtering scale $R$ equal to the Lagrangian halo radius.
This result confirms that, 
{\it when the purpose is to compute statistical quantities such as power spectra, etc}, 
the coupled-fluid approximation requires a modification of the momentum conservation equation for 
halos  \cite{v1} as   the  force felt by the halos  is  biased relative to the force acting on a 
random DM particle.

The EBE  provides a  way of describing the evolution equation for biased tracers. 
As first suggested in Ref. \cite{v2}, it encodes the  statistical velocity bias which originates 
from the fluctuations  around the ensemble average of the gravitational force acting on the halo 
and it predicts the right value (\ref{biasv}) observed at the linear level. 
Therefore, the EBE explains at  more fundamental level why and how one needs to modify the momentum 
fluid equation  for the coupled-fluids of halos and DM, as suggested by Ref. \cite{v1}. 
In addition, it allows us to compute the halo velocity bias at higher orders and to derive additional 
stochastic components that describe in a {\it mean-field} way (i.e. at the statistical level) how 
biasing influence the dynamics of halo centers.
 
The paper is organised as follows: in Sec.~\S\ref{sec:boltzmann equation}, we motivate our approach 
and introduce the various averaging we perform throughout the calculation.
In Sec.~\S\ref{sec:boltzmann eq LHS} we perform a detailed computation of the left-hand side 
of the EBE adopting  a general method based on the path integral approach for computing conditional 
averages. 
Section~\S\ref{sec:boltzmann eq RHS} is devoted to a similar discussion regarding the right-hand 
side of the EBE.
We then conclude with some final remarks in Sec.~\S\ref{sec:conclusions}.

\section{From the Klimontovich-Dupree equation to the Boltzmann equation}
\label{sec:boltzmann equation}

%
\hskip 0.8cm
Let us first explain in more details what we mean by halo ``mean-field''.
In a local bias approximation (e.g. \cite{Fry:1992vr}), one can express density fluctuations in the halo mean-field as the series expansion 
\begin{equation}
\delta_\text{h}(\vr,t)=b_1 \delta_R(\vr,t) + \frac{1}{2} b_2 \big(\delta_R^2(\vr,t)-\sigma_0^2(t)\big) 
+ \dots
\end{equation}
where the filtering scale $R$ is proportional to the halo mass \cite{v0}, the bias parameters $b_1$,
$b_2$ etc. are {\it ensemble average} whereas $\delta_R(\vr,t)$ is {\it one} particular realisation of 
density fluctuations in the large scale structure at cosmic time $t$. The filtering reflects the fact
that we are not interested in the internal structure of the halos, only in their center-of-mass 
position and kinematics. From the knowledge of this halo mean-field, we can compute halo clustering 
statistics, such as the halo 2-point correlation, as
\begin{equation}
\xi_\text{h}(\vx_1-\vx_2,t) = b_1^2 \big\langle \delta_R(\vx_1,t) \delta_R(\vx_2,t)\big\rangle + \dots
\end{equation}
where $\langle\dots\rangle$ is an average over realisations of the large scale structure.

%

Our goal is to write down an EBE  for the single particle  phase space density $f_\text{h}$ of halo 
centers, such that its zeroth moment yields $\delta_\text{h}(\vr,t)$. 
This EBE must be  different from the Boltzmann equation of DM: the latter is recovered  only if the 
halos are statistically unbiased relative to the coarse-grained DM density field (which they are not).
Our starting point is  the Klimontovich density \cite{klim}
\be
f_{\rm K}(\vr,\vp,t)=\sum_i\delta_{\rm D}\left[\vr-\vr_i(t)\right]\delta_{\rm D}\left[\vp-\vp_i(t)\right],
\ee
which is the single particle phase space density for DM. We have assumed that all the elementary DM particles have equal mass $m$, and we normalize our units so that $m=1$. 

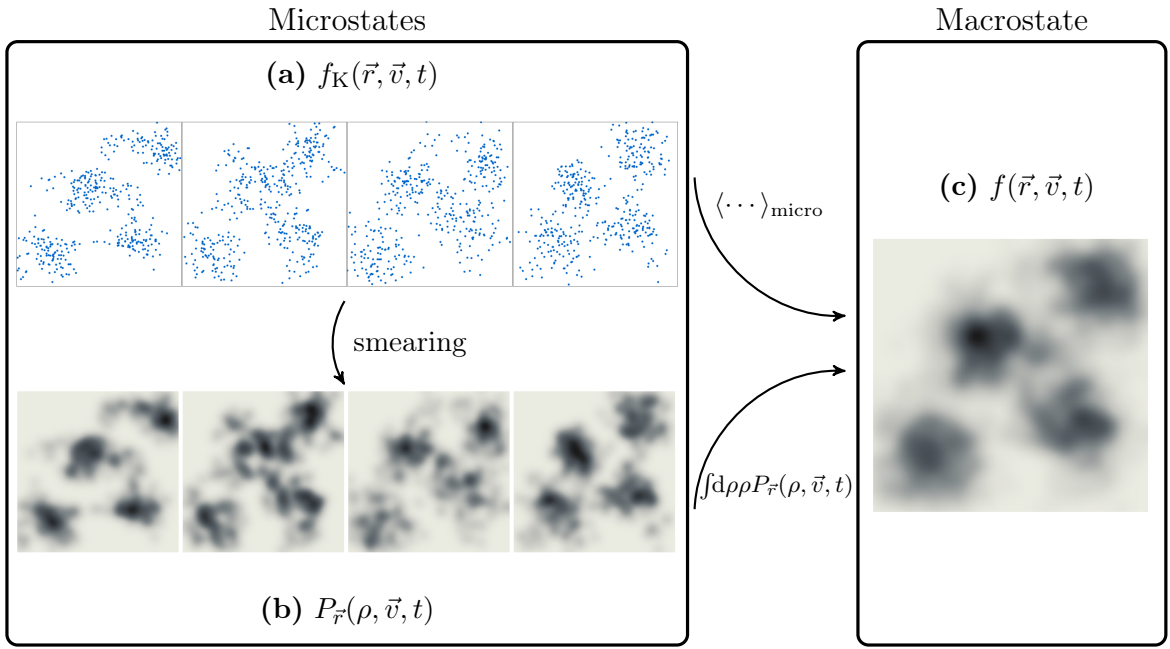
\begin{figure}[h!]
\centering
\begin{tikzpicture}[node distance=1cm, auto,]
\node[rectangle, rounded corners,draw=black, very thick,
text width=\microboxwidth,minimum height=\boxheight,
align=center] (microbox) {
  { \textbf{(a)} $f_\text{K}(\vec r,\vec v,t)$}\\
  \vspace{1em}		{\pgfuseimage{micro}}\\
  \vspace{1em}		{\hspace{4em} smearing}\\
  \vspace{1em}		{\pgfuseimage{smearedSmall}}\\
  \vspace{1em}		{ \textbf{(b)} $P_{\vec r}(\rho,\vec v,t)$}
};
\node[above=0.cm of microbox] (microtitle) {{\large {Microstates}}};
\node[above=-3.4 of microbox] (c1){};
\node[above=-4.9cm of microbox] (c2){};
\node[] {} (c1.south) edge[freccia,bend right=30](c2.north);
\node[rectangle, rounded corners,draw=black, very thick,
text width=\macroboxwidth,minimum height=\boxheight,
align=center, inner sep=5pt,right=2.2cm of microbox] (macrobox) {
  { \textbf{(c)} $f(\vec r,\vec v,t)$}\\
  \vspace{1em}		{\pgfuseimage{macroSmall}} 
};
\node[above=0.cm of macrobox] (macrotitle) {{\large {Macrostate}}};
\node[right=-0.2cm of microbox](r){};
\node[above=2cm of r](rUp){};
\node[below=2cm of r](rDown){};
\node[left=-0.2cm of macrobox](l){};
\node[above=0.1cm of l](lUp){};
\node[below=0.1cm of l](lDown){};
\node[] {} (rUp.east) edge[freccia,bend right=45](lUp.west);
\node[] {} (rDown.east) edge[freccia,bend left=45](lDown.west);
\node[below right=0.cm of rUp]{{ \small $\langle \cdots\rangle_\text{micro}$}};
\node[above right=-0.1cm of rDown]{{\footnotesize $\int \hspace{-3pt}\mathrm d\rho \rho P_{\vec r}(\rho,\vec v,t)$}};
\end{tikzpicture}
\caption{A pictorial exemplification of the ensemble average described in the text. We consider all the configurations (\textit{microstates}) for the distribution of DM particles that yield to indistinguishable large scale structure configurations (a given \textit{macrostate}). 
\textbf{(a)} The distribution of DM particles is described in the phase space by the Klimontovich density $f_\text K(\vr,\vv,t)$. Four different realizations are shown. 
\textbf{(b)} After a suitable smearing (the convolution with a window function) we can describe these microstates in terms of a smooth density $\rho$. In any point $\vr$, $\rho$ can vary in a range of values among the microstates: this stocasticity is encoded in the probability distribution $P_\vr (\rho,\vv,t)$. 
\textbf{(c)} The macrostate is described by the phase space density $f(\vr,\vv,t)$, which follows from the ensemble of microstates through the ensemble average $\langle f_\text K (\vr,\vv,t)\rangle_\text{micro}$, or equivalently by computing the expected value of $\rho$ with respect to the probability density $P_\vr (\rho,\vv,t)$. 
}
\label{fig:micro}
\end{figure}

The corresponding equation of motion for such a phase space density is the Klimontovich-Dupree equation
\be
\label{kd}
\frac{\partial f_{\rm K}}{\partial t}+\vec{v}\cdot \frac{\partial f_{\rm K}}{\partial\vr}
-\vec{\nabla}\Phi_{\rm K}\cdot\frac{\partial f_{\rm K}}{\partial\vp}=0,
\ee
where 
\be\label{phik}
\vec{\nabla}\Phi_{\rm K}(\vr,t)=G_{\rm N}\int\d^3r'\d^3v'f_{\rm K}(\vrp,\vvp,t)\frac{(\vr-\vrp)}{\left|\vr-\vrp\right|^3}
\ee
is the corresponding gravitational force. 
At this stage we have not learned much, as the Klimontovich density encodes the trajectories of all the particles in the system, which is much more information than we can handle and want to keep track of (in particular, we are not interested in the internal properties of virialized structures). Moreover, eq.~\eqref{kd} describes one particular realization of a universe, which may not resemble our own Universe. 
Therefore, we go one step further and average over realizations of universes that have the same large-scale structure (i.e. similar coarse-grained properties). In this way we can construct macrostates by averaging over a statistical ensemble of microstates with similar phase space density in (small) volumes containing a sufficient amount of particles.\\ 
Let us explain in more detail how this ensemble average is performed with the help of fig.~\ref{fig:micro}.

The Klimontovich density describes a set of points in the phase space. Let us focus on a small box in the configuration space. 
In the part \textbf{(a)} of fig.~\ref{fig:micro} we show four different realizations of this box, which yield the same LSS. Their ensemble average, which we denote by $\langle\cdots \rangle_{\rm micro}$, gives the mean value $f(\vr,\vv,t)$:
\be
\label{f average}
\Big<f_{\rm K}(\vr,\vp,t)\Big>_{\rm micro}=
\Big<\sum_i\delta_{\rm D}\left[\vr-\vr_i(t)\right]\delta_{\rm D}\left[\vp-\vp_i(t)\right]\Big>_{\rm micro}=f(\vr,\vp,t)\;,
\ee

An alternative way of understanding this average is to smear out the discrete distribution of points into a continuous density with an associated probability $P_{\vr}(\rho, \vv,t)$, where $\vr$ is any position inside the box, as represented in the part \textbf{(b)} of fig.~\ref{fig:micro}. The range of values assumed by $\rho$ over the ensemble defines the probability density $P_\vr(\rho,\vv,t)$ for having a velocity $\vv$ and a density $\rho$ in a small volume centered at position $\vr$, provided that the LSS is the one fixed in \textbf{(c)} in fig.~\ref{fig:micro}. \footnote{Note that this smearing should not be confused with the filtering procedure introduced in section \ref{sec:boltzmann eq LHS}. The latter defines a UV cutoff in momentum space denoted by $\Lambda$, while the former represents an average over microstates in small, real space volumes of size $\ell$, with the requirement that $\ell \ll 1/\Lambda$.}
In this case, $f(\vr,\vv,t)$ is given by the mean value of $\rho$ with respect to the probability distribution $P_\vr(\rho,\vv,t)$:
 \be
 \label{f integral rho}
 f(\vr,\vp,t)=\int_0^\infty\,\d\rho\,\rho\, P_{\vr}(\rho,\vp,t)=\int_0^\infty\,\d\rho\,\rho\, P_{\vr}(\rho|\vp,t)\,P_{\vr}(\vp,t)=\langle\rho|\vp\rangle\,P_{\vr}(\vp,t).
 \ee
We will drop the subscript $\vr$ here and henceforth, as it is clear that $P(\rho,\vp,t)$ is defined at a given spatial location. 

The distribution function $f(\vr,\vp,t)$ that characterizes the macrostates defines, upon taking moments of the velocity $\vv$, the overdensity field, the bulk velocity, etc. of a given macrostate. For instance, the matter overdensity field $\delta(\vr,t)$ is the zeroth moment
\be
\bar{\rho}\bigl(1+\delta(\vr,t)\bigr) = \int\d^3v\, f(\vr,\vp,t).
\ee
This distribution function satisfies the equation
\be
\label{akd}
\frac{\partial f}{\partial t}+\vec{v}\cdot \frac{\partial f}{\partial\vr}
-\vec{\nabla}\Phi\cdot\frac{\partial f}{\partial\vp}=\vec\nabla_{\vp} \cdot\vec F,
\ee
where
\be\label{phi}
\vec{\nabla}\Phi=G_{\rm N}\int\d^3r'\d^3v'f(\vrp,\vvp,t)\frac{(\vr-\vrp)}{\left|\vr-\vrp\right|^3}
\ee
and 
\begin{align}
\label{force}
\vec{F}(\vr,\vp,t) 
&={\rm Cov}\left[\vec\nabla\Phi_{\rm K}(\vr,t),f_{\rm K}(\vr,\vp,t)\right]_{\rm micro} \nonumber \\
&=  G_{\rm N}\int\d^3r'\, \d^3 v'\, \frac{(\vr-\vrp)}{\left|\vr-\vrp\right|^3} f_{\rm 2c}(\vr,\vp,\vrp,\vvp,t)
\;,
\end{align}
where
\be
\label{irr}
f_{\rm 2c}(\vr,\vp,\vrp,\vvp,t)=f_{\rm 2}(\vr,\vp,\vrp,\vvp,t)-f(\vr,\vp,t)f(\vrp,\vvp,t)
\ee
is the irreducible   two-particle correlation function. 
 
Equation (\ref{akd}) is our starting point towards a derivation of the EBE  for halos, as it enables us to apply constraints involving $\delta$ etc.~to define biased tracers of the large scale structure.

\section{Halos as extrema of the smoothed density field}
\label{sec:halos}
\hskip 0.8cm
In order to write down an EBE for the halo phase space density $f_\text{h}(\vr,\vv,t)$, we follow the evolution over cosmic time, and in Eulerian space, of the so-called proto-halos until their virialization. The proto-halos are the progenitors of isolated DM halos, i.e. DM halos that are not contained in any larger halo.
While  their shape and topology change as a function of time (smaller substructures gradually merge to form the final halo), their centre of mass moves along a well-defined trajectory determined by the surrounding mass density field.
Hence, unlike virialized halos that experience merging, by construction proto-halos
always preserve their identity. Their total number is, therefore, conserved over time, such that we can write a continuity and Euler equation for their number density and velocity, respectively.

We will consider a specific model for the halo biasing in order to derive explicit expressions for the amplitude and scale-dependence of the effective corrections to the halo EBE.
Namely, we will hereafter designate by halos (or peaks) those clustered objects that are located at the points where $\vec{\nabla}\delta\ww(\vr)=\vec{0}$ and where the smoothed density contrast satisfies $\delta\ww(\vr)=\nu\sigma_0$ (we will occasionally write $\delta_\text{pk}\equiv\nu\sigma_0$).
The subscript $R$ represents quantities that have been smoothed on the (comoving) size $R$ of a halo. For instance,
\begin{equation}
\delta_R(\vr,t)= \int \d^3 r' \delta(\vrp,t)\, W\!\left(\frac 1R |\vr-\vrp|\right)\;,
\end{equation} 
where $W(x)$ a spherically symmetric smoothing kernel. 
Moreover, $\nu$ is the peak height and $\sigma_0\equiv \sigma_0(t)$ is the variance of the DM field at time $t$ in a given region of size $R$. In general, the spectral moments of the smoothed fields are given by
\be
\label{sigma def}
\sigma_j^2=\int\frac{\d^3 k}{(2\pi)^3}\, k^{2j} 
P_{\rm }(k)\tilde{W}^2(kR),
\ee
where $\tilde{W}(x)$ is the Fourier transform of $W(x)$,  $P(k)=\big\langle |\delta(\vk,t)|^2\big\rangle_\text{macro}$ is the matter power spectrum, the ensemble average is over the macrostates and the time dependence is implicit. We will occasionally refer to those conditions as the \textit{peak constraint}, although we enforce the extremum condition solely. However, notice that, in the high peak limit $\nu\gg 1$, nearly all extrema are maxima of $\delta\ww$.

Last but not least importantly, the filtering of the density field reflects the fact that halos are extended objects. In our model, we will take this into account upon assuming that halo centers feel a force which is a smoothed version of the force acting on dark matter particles.

\section{The halo Boltzmann equation: the left-hand side}
\label{sec:boltzmann eq LHS}
\hskip 0.8cm

Our goal is to specialize the Boltzmann equation \eqref{akd} from the DM phase space density to the peak phase space density so as to describe the evolution of halos. For this purpose, $\fpk\rvt$ will designate the peak phase space density -- i.e. the number density of peaks of velocity $\vv$ at the position $\vr$ and time $t$ --
 such that its zeroth moment corresponds to the halo mean-field $\delta_\text{h}(\vr,t)$.
The density $\fpk$ can be written in terms of conditioned probabilities, as we will show in eq.~\eqref{fpk} of section \ref{sec:boltzmann eq RHS}. 
The peak constraint in $\vr$ also affects the computation of the force on the right-hand side of Boltzmann equation, 
introducing conditioned probabilities. 
As we will discuss in section \ref{sec:boltzmann eq RHS}, this side of the Boltzmann equation eventually vanishes.

Let us focus now on the left-hand side of \eqref{akd}. There, the following operator appears: 
\be
\label{composite operator}
\vec{\nabla}\Phi\rt\cdot\dvfrvt \,,
\ee
which is the product of two operators evaluated at the same point, that is, a \textit{composite operator}. 
It is well-known that, owing to the  local product of fields making up the composite operator, new ultraviolet divergences~\footnote{In the picture in which $\Lambda$ represents the smoothing scale at which the peak is defined, $\Lambda$ never truly diverges in CDM cosmologies since it is a physical scale. However, in the formalism we introduce here we wish to appeal to the reader's intuition of the QFT renormalisation scheme. Within this context, our terminology is similar to \cite{Assassi:2014fva}.} appear and depend on the ultraviolet cut-off $\Lambda\sim\mathcal{O}(1/R)$. 
The appearance of these divergences  requires the introduction of new counterterms. 
The physical reason why the term~\eqref{composite operator} changes in the equation for halos is that the force felt by peaks will be \textit{statistically} biased with respect to the DM case. The technical way by which we can reconstruct the form of $\vnabphi\cdot\dvf$ for the halo Boltzmann equation is through renormalization. 

Renormalized operators can be defined and generically expressed as linear combinations of all the bare operators of equal or lower canonical dimensionality (see, for instance, \cite{collins})
\be
\label{renorm definition [O]}
\left[{\cal O}_i\rt\right]=\sum_{j} Z_{ij}(\Lambda)\,{\cal O}_j\rt.
\ee
Therefore, any composite operator can be expressed as a sum of operators allowed by the symmetries of the problem at hand.
The renormalized operator $[\vnabphi \cdot \dvf]$ being a scalar field, it can mix with all the operators allowed by the extended Galilean symmetry \cite{con} and must remain a function of the velocity $\vv$ and preserve the derivative $\partial_{v_i}$. 
We can classify the operators $\mathcal O_i$ that mix with $\vnabphi \cdot \dvf$ according to their number of derivatives in $\vv$:
\begin{equation}
\label{composite op renorm}
\left[\vnabphi \cdot \dvf \right]= \vnabphi \cdot \dvf  +
	\underset{\text{leading order in }\partial_\vv}{\underbrace{Z_{01}\,\vnabdelta\cdot \dvf + \dots}} +
	\underset{\text{higher order in }\partial_{\vv}}{\underbrace{\tilde{Z}\vnabphi \cdot \partial_\vv^3 f + \dots}} \,.
\end{equation}

\begin{figure}[h!]
\centering
\begin{tikzpicture}[node distance=1cm, auto,]
\node[operator] (o0) {$\mathcal O_0=\vnabphi \cdot\dvf$};
\node[operator, right=1.5cm of o0] (o1) {$\mathcal O_1=\vnabdelta \cdot\dvf$};
\node[operator, below=1.cm of o0] (o2) {$\mathcal O_2=\Otwo \cdot\dvf$\\
$\mathcal O_3=\Othree \cdot\dvf$};
\node[operator,right=1.5cm of o1] (r) {$\cdots$};
\node[operator,right=1.5cm of o2] (c) {$\cdots$};
\node[operator,below=1.cm of o2] (l) {$\cdots$};
\node[right=-1mm of o0,align=center]{{\footnotesize add 2 der.}\\\vspace*{1em}};
\node[below=0.25cm of o0]{{\footnotesize add a field}\hspace*{5em}};
\node[] {} (o0.east) edge[freccia,bend left=0](o1.west);
\node[] {} (o0.east) edge[freccia,bend left=0](o1.west);
\node[] {} (o1.east) edge[freccia,bend left=0](r.west);
\node[] {} (o2.east) edge[freccia,bend left=0](c.west);
\node[] {} (o0.south) edge[freccia,bend right=0](o2.north);
\node[] {} (o1.south) edge[freccia,bend right=0](c.north);
\node[] {} (o2.south) edge[freccia,bend right=0](l.north);
\end{tikzpicture}
\caption{Operators allowed by rotational invariance that can mix with $\vnabphi\cdot\dvf$, classified according to their number of fields and derivatives. 
}
\label{fig: operators}
\end{figure}
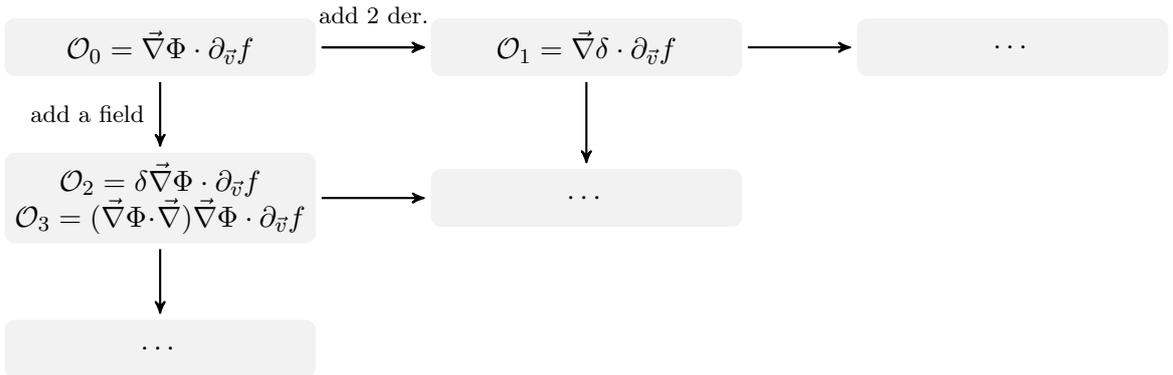

Once we have accomplished this classification, we select the leading operators according to the number of derivatives (we are considering large scales) and fields contained in them 
(adding additional fields brings to a result including more power spectra). 
The fields we are dealing with are $\Phi$ (which can only appear with derivatives, since the gravitational potential is not observable and  does not directly affect the dynamics) and the density contrast $\delta=\nabla^2\Phi/\alpha $, where $\alpha=3 H^2 \Omega_m/2$. We can then organise the $\mathcal O_i$ according to the number of extra couples of derivatives (in order to build a vector, we must add two derivatives to $\vnabphi$ and contract two indices) and extra fields appearing in them. The result is shown in fig.~\ref{fig: operators}, where moving to the right means adding two derivatives, and moving downwards means adding a field. The appearance of derivative operators makes clear that the statistical bias manifests itself only through $k$-dependent corrections and, therefore, vanishes in the limit of small momenta.

The renormalization procedure requires a  suitable prescription. For our purpose, it is convenient to adopt the following renormalization condition: the correlators of the renormalized quantities should reproduce the correlators at the peak,
\be
\label{renorm prescription}
\Big<\ren{\mathcal O_1\rt}\cdot\ren{\mathcal O_2\rt}\cdots \ren{\mathcal O_n\rt}\Big>
=\Big<{\cal O}_1\rt\cdot{\cal O}_2\rt\cdots{\cal O}_n\rt\Big>_{\rm pk}.
\ee
Here and henceforth, the notation $\langle \cdots\rangle_\text{pk}$ will designate conditional averages at the peak position.
In our problem, it is precisely the  renormalization of the operator $\vnabphi\rt\cdot \dvfrvt$ that leads to differences in the clustering properties of halos and DM, as we will now see. Notice that, since we want to follow the evolution over cosmic time of the so-called proto-halos until their virialization, the renormalization condition has to be imposed at all times.

\subsection{Renormalization  at the linear order}
\hskip 0.8cm In this subsection, we study the renormalization of the composite operator $\vnabphi\cdot \dvfrvt$ at the linear order in perturbation theory.

We begin with the computation of the average of $\vnabphi\rt \cdot \dvfrvt$ at the peak position:
\be
\label{average composite op}
\Big<\vnabphi\rt \cdot \dvfrvt \Big>_{\rm pk}=\dvf_{\rm pk}\rvt \cdot \Big<\vnabphi\rt \Big>_{\rm pk}\;.
\ee
This amounts to computing the correlators using the appropriate conditional probabilities.

Let us first operate at the linear level in perturbation theory such that, at early times, the DM  density $\delta_{\rm }$ and  the gravitational force $\vec\nabla\Phi$  are Gaussian variables. We can then apply the theorem stated, for instance,  in Refs.~\cite{bbks,bema}, which ensures that the conditional probability of zero-mean Gaussian variables $X_A$ and $X_B$ is itself a Gaussian variable with mean
\be
\Big<X_B\Big|X_A\Big> \equiv \frac{\Big<X_B \otimes X_A\Big>}{\Big<X_A \otimes X_A \Big>}X_A
\ee
and covariance matrix
\be
\label{covariance constrained}
{\rm C}\Big(X_B,X_B\Big) \equiv \Big<X_B \otimes X_B \Big> - \frac{\Big<X_B \otimes X_A\Big>}{\Big<X_A \otimes X_A \Big>}\Big<X_A \otimes X_B \Big>.
\ee
In our case, we identify $X_A$ with $\vec\nabla\delta\ww$ and $X_B$ with $\vec\nabla\Phi\ww$. As explained above, the gravitational force acting on the halo centers is smoothed to reflect the finite extent of the halos.
Since $\langle \vnabphi\otimes \delta \rangle=\vec 0$ because of rotational invariance, we can simply compute $\langle \vnabphi \rangle_\text{pk}$ as $\langle\vnabphi\ww|\vnabdelta\ww=\vec 0\rangle$.The mean shift of $\vnabphi$ at the peak position is given by
\begin{equation}
\label{average grad phi}
\begin{aligned}
\Big<\vnabphi\Big>_{\rm pk}&=
	\Big<\vnabphi\ww\Big |\vnabdelta\ww \Big> = \frac{\Big<\vec{\nabla}\Phi\ww\cdot\vec{\nabla}\delta\ww\Big>}{\Big<\left(\vnabdelta\ww \right)^2\Big>}\,\left.\vec{\nabla}\delta\ww \right|_{\rm pk}
=-\alpha \frac{\Big<\delta^2\ww \Big>}{\Big<\left(\vec{\nabla}\delta\ww \right)^2\Big>}\,\left.\vec{\nabla}\delta\ww \right|_{\rm pk} \\ 
& = -\alpha\,\frac{\sigma_0^2(\Lambda)}{\sigma_1^2(\Lambda)}\,\left.\vec{\nabla}\delta\ww \right|_{\rm pk},
\end{aligned}
\end{equation}
where we have highlighted the dependence of the spectral moments $\sigma_j$ on the UV cut-off $\Lambda$. Since at the peak $\vec{\nabla}\delta\ww$  is vanishing, no renormalization is needed.
We will hereafter drop the subscript $R$ whenever $\vnabphi$ and $\delta$ appear because we will always refer to peaks, hence the smoothing is understood.
Notice also that, while the renormalisation condition is imposed at all times, the scale $\Lambda$ is fixed at initial time and does not evolve. In reality, $\Lambda$, which is related to the smoothing scale $R$ that define the Lagrangian extent of a halo, likely evolves with time (see e.g. \cite{Baldauf:2013hka}). For simplicity however, we will ignore this complication in the present work.
 
We wish now to compute the average of $\vec{\nabla}\Phi\rt \cdot \dvfrvt$ with itself at the peak position. On applying Wick's theorem, the only non-vanishing term that remains is
\begin{multline}
\Big<\left(\vnabphi\rt \cdot \dvfrvt\right)^2\Big>_{\rm pk}=
\dvf_{\rm pk}\rvt \cdot \dvf_{\rm pk}\rvt\Big <\vec{\nabla}\Phi\rt \cdot\vec{\nabla}\Phi\rt\Big>_{\rm pk},
\end{multline}
where we used the fact that $\Big <\nabla_i\Phi\rt \cdot\nabla_j\Phi\rt\Big> \propto \delta_{ij}$.
This requires the knowledge of the covariance matrix for $\vnabphi$ given the constraint at the peak. At the linear order the unconstrained correlators needed to evaluate eq. ~\eqref{covariance constrained} read 
\begin{equation}
\begin{aligned}
\Big<\vec{\nabla}\Phi\otimes\vec{\nabla}\Phi\Big>&=\frac{\alpha^2}{3}\sigma^2_{-1}(\Lambda)\,\mathbb{1}_{3\times 3},\\
\Big<\vec{\nabla}\Phi\otimes\vec{\nabla}\delta_{\rm }\Big>&=-\frac{\alpha}{3}\sigma^2_{0}(\Lambda)\,\mathbb{1}_{3\times 3},\\
\Big<\vec{\nabla}\delta_{\rm }\otimes\vec{\nabla}\Phi\Big>&=-\frac{\alpha}{3}\sigma^2_{0}(\Lambda)\,\mathbb{1}_{3\times 3},\\
\Big<\vec{\nabla}\delta_{\rm }\otimes\vec{\nabla}\delta_{\rm }\Big>&=\frac{1}{3}\sigma^2_{1}(\Lambda)\, \mathbb{1}_{3\times 3}.
\end{aligned}
\end{equation}

The covariance matrix for $\vnabphi$ given the peak constraint is then equal to
\be
\label{cov}
{\rm C}\left(\vec{\nabla}\Phi,\vec{\nabla}\Phi\right)_\text{pk}=\frac{\alpha^2}{3}\left(\sigma^2_{-1}(\Lambda)-\frac{\sigma^4_{0}(\Lambda)}{\sigma^2_{1}(\Lambda)}\right)\,\mathbb{1}_{3\times 3},
\ee
and
\be
\label{cov1}
\Big <\vec{\nabla}\Phi\rt \cdot \vec{\nabla}\Phi\rt\Big>_{\rm pk}={\rm Tr}\, C\left(\vec{\nabla}\Phi,\vec{\nabla}\Phi\right)=\alpha^2\left(\sigma^2_{-1}(\Lambda)-\frac{\sigma^4_{0}(\Lambda)}{\sigma^2_{1}(\Lambda)}\right).
\ee
We recall now the prescription~\eqref{renorm prescription}: at the lowest order,  $[\vnabphi \cdot \dvf]$ contains only the operators $\mathcal O_0$ and $\mathcal O_1$ (see fig.~\ref{fig: operators}). Hence, we learn that the renormalized force felt by the halos at the linear order reads
\be
\label{bb}
\left[\vec{\nabla}\Phi\rt\cdot \dvfrvt\right]^{\rm first}=\left( \vec{\nabla}\Phi\rt+\alpha\frac{\sigma_0^2(\Lambda)}{\sigma_1^2(\Lambda)}\vec{\nabla}\delta_{\rm }\rt\right) \cdot \dvfrvt.
\ee
Although the mean shift of the gravitational force vanishes at the peak and thus there is no extra force at the peak-by-peak level, the gravitational force at the peak receives a correction when the statistical ensemble average is taken, as exemplified by eq.~\eqref{cov1}.
The extra effective force felt by the halos is purely of statistical origin and, therefore, does not violate the Equivalence Principle. 
Furthermore, the time-dependence of $Z_{01}$ is that of $\alpha \sigma_0^2/\sigma_1^2$, that is, the amplitude of the resulting $k^2$-correction in Fourier space {\it does not depend on time}. Therefore, this gravity bias does not decay with time, in agreement with \cite{dcss,v1}.
Note that the same result eq.~\eqref{bb} was found in Refs. \cite{bao,v0,v1,v2}, but our Boltzmann approach combined with a renormalization procedure puts it into a different perspective.


\subsection{Renormalization at higher order with the path integral technique}
\label{sec:renorm ho}

\hspace{0.8cm} In this subsection, we will compute the renormalized operator $[\vnabphi\rt\cdot\dvfrvt]$ at the second order in perturbation theory. We will compute the correlators beyond the Gaussian approximation with the path integral technique (details can be found in appendix~\ref{sec:appendix A}). We will restrict ourselves to the four operators $\mathcal O_0,\,\mathcal O_1,\,\mathcal O_2,\,\mathcal O_3$ defined in fig.~\ref{fig: operators}.

Let us give a closer look to the renormalization prescription \eqref{renorm prescription}. We begin by observing that the expectation values of single operator $\langle\mathcal O_i\rangle$ are not helpful, because they are all proportional to expectation values of vector fields and thus vanish, as in~\eqref{average composite op} and \eqref{average grad phi}. Therefore, in order to determine the coefficients $Z_{ij}$ of \eqref{renorm definition [O]}, we write down the system given by the expectation values of products of two operators:
\begin{equation}
\label{renorm system}
\sum_{l,m} Z_{il} Z_{jm}\corr{l}{m} =\corrpk{i}{j}\,.
\end{equation}

We must now specify how we define the leading and subleading orders of our computation. On the left-hand side of eq.~\eqref{renorm system}, operating at second order the correlation functions $\corr{i}{j}$ contain at most two power spectra. Therefore, at leading order we compute the correlators by including only their Gaussian components and obtaining one power spectrum in the result; the next-to-leading order includes up to two power spectra. 
On the right-hand side, the leading order of the computation is obtained by considering again only the Gaussian component of the fields. 
When computing the correlators $\corrpk{i}{j}$, the linear order is equivalent to consider only the leading order in $\nu\sigma_0$.

The results for the correlators without the peak constraint are\footnote{see Appendix~\ref{sec:appendix B} for details about the calculation of these correlators, and for the definition of the finite quantities $\mathcal K_1$, $\mathcal K_2$)}
\begin{eqnarray}
  & \text{(leading order)} & \text{(next-to-leading order)}\nonumber\\
\corr{0}{0} = &\displaystyle \frac{\alpha^2}{3}\sigma^2_{-1}\delta_{ij}\\
\corr{0}{1} = & \displaystyle -\frac{\alpha}{3}\sigma^2_{0}\delta_{ij}\\
\corr{0}{2} = & & \frac{\alpha^2}{3}\mathcal K_1 \delta_{ij}\\
\corr{0}{3} = & & -\frac{\alpha^3}{6} \mathcal K_1 \delta_{ij} \\
\corr{1}{1} = & \displaystyle \frac{1}{3}\sigma^2_{1}\delta_{ij}\\
\corr{1}{2} = & & -\frac{\alpha}{3}\frac{17}{7}\sigma^4_{0}\delta_{ij}\\
\corr{1}{3} = & & - \frac{\alpha^2}{3} \mathcal K_2\, \delta_{ij} \\
\end{eqnarray}
where we are dropping the factors of $\partial_\vv f$ to simplify the notation. The correlators calculated with the peak constraint read 
\begin{eqnarray}
\corrpk{0}{0} = &
  \displaystyle  \frac{\alpha^2}{3}\delta_{ij}\biggl[\left(\sigma^2_{-1} - \frac{\sigma^4_0}{\sigma^2_1}\right) & +
	(\nu\sigma_0)\left(\mathcal K_1\frac{1}{\sigma_0^2} -\frac{20}{21} \frac{\sigma_0^4}{\sigma_1^2}\right) \biggr] \\
\corrpk{0}{2}  = & &
  \frac{\alpha^2}{3} (\nu\sigma_0)\delta_{ij}	\left(\sigma^2_{-1} - \frac{\sigma^4_0}{\sigma^2_1}\right) \\
\corrpk{0}{3} = && 
   \frac{\alpha^3}{9} (\nu\sigma_0)\delta_{ij} \left(\sigma^2_{-1} - \frac{\sigma^4_0}{\sigma^2_1}\right)
\end{eqnarray}
We also notice that our renormalization prescription \eqref{renorm prescription} enforces $\corrpk{1}{i}=0\,\, \forall i$. This corresponds to a subsystem of \eqref{renorm system} (when $i=1$) whose right-hand side is 0, the solution of which is $Z_{1i}=0$ or $[\vnabdelta\cdot\dvf]=0$. 
This is a consequence of our renormalization prescription: once we impose that  at the position $\vr$ there is a peak defined by $\vnabdelta(\vr)=\vec 0$, the operator $\vnabdelta$ after renormalization cannot be anything but zero.

We are now free to  impose the condition $\ren{\mathcal O_i}=\mathcal O_i +\dots$, i.e.  the diagonal entries satisfy the condition $Z_{ii}=1$ for $i\neq 1$.
%
Let us first discuss the system for the renormalization of the three operators $\{ \mathcal O_0,\mathcal O_1,\mathcal O_2\}$. We begin by inspecting the equation $\corrpk{0}{0} = Z_{0i}Z_{0j}\corr{i}{j}$, which contains all (and only) the coefficients $Z_{0i}$. We know that
\[ Z_{00}=1, \qquad 
Z_{01} = \alpha \frac{\sigma_0^2}{\sigma_1^2}+\mathcal O(\nu\sigma_0), \qquad  
Z_{02} = \mathcal O(\nu\sigma_0)
\,. \]
We should impose the equality $\corrpk{0}{0}=Z_{0i}Z_{0j}\corr{i}{j}$ separately between the leading order terms and the next-to-leading order ones. We  find that 
\[ Z_{01} = \alpha \frac{\sigma_0^2}{\sigma_1^2}+\mathcal O(\nu\sigma_0),\quad Z_{02}=(\nu\sigma_0)\left[\frac{1}{2 \sigma_0^2} + \frac{31 \sigma_0^4}{6 \left(-17\sigma_0^6 + 7 \mathcal K_1 \sigma_1^2\right)}\right]+\mathcal O\left((\nu\sigma_0)^2\right)\,.\]
We now look at the equations $\corrpk{0}{2}=Z_{0i}Z_{2j}\corr{i}{j}$ and $\corrpk{2}{2}=Z_{2i}Z_{2j}\corr{i}{j}$, which include the unknowns $Z_{20}$ and $Z_{21}$. These two equations, once we select the terms consistently with their order (for example, the connected contribution of $\corr{2}{2}=\langle \delta\vnabphi \delta\vnabphi  \rangle$ contains 3 power spectra and should not be put together with second order quantities) we get
\begin{align*}
Z_{20} =&\frac{\mathcal{K}_1 \sigma _1^2-\tfrac{17}{7} \sigma _0^6}{(\sigma _0^4-\sigma _{-1}^2 \sigma _1^2 )} + \underset{=\delta_\text{pk} Z_{00}}{\underbrace{\left(\nu\sigma_0\right)}} \,, \\
Z_{21} =&\alpha \left[ \sigma_0^2 \frac{\mathcal{K}_1 -\tfrac{17}{7} \sigma _0^2\sigma_{-1}^2}{(\sigma _0^4-\sigma _{-1}^2 \sigma _1^2 )} 
\pm \frac{1}{\sigma_1^2}\sqrt{-\frac{\left(\mathcal{K}_1 \sigma_1^2-\tfrac{17}{7} \sigma _0^6\right)^2}{(\sigma _0^4-\sigma _{-1}^2 \sigma _1^2 )}
+\sigma_0^2 \left[\left(\tfrac{17}{7}\right)^2 \sigma_0^6- \sigma_{-1}^2 \sigma_1^2\right]}
\right] +
\underset{=\delta_\text{pk} Z_{01}}{\underbrace{\alpha \frac{\sigma_0^2}{\sigma_1^2}\left(\nu\sigma_0\right)}} \,.
\end{align*}
This concludes the determination of the renormalization coefficients at the next-to-leading order for the set of operators $\{ \mathcal O_0,  \mathcal O_1,  \mathcal O_2\}$, keeping up to two power spectra in the (connected) correlators. 
We could now include the operator $\mathcal O_3$ in the discussion, since it contains the same number of fields $\Phi$ and derivatives acting on them as $\mathcal O_2$. 
But in this way we should compute correlation functions as $\corrpk{3}{3}$, whose connected part contains three power spectra, and thus goes beyond the perturbative order we are considering. 
The difference with respect to the operator $\mathcal O_2$ is that the correlator $\corrpk{2}{2}$, with the constraint $\delta=\nu\sigma_0$, gives $(\nu\sigma_0)^2$ times a correlator with one power spectrum. 
Now, $\sigma_0^2$ is of the same order of a power spectrum in a perturbative expansion (see eq. ~\ref{sigma def}), thus $\corrpk{2}{2}$ is of a lower order with respect to $\corrpk{3}{3}$. Therefore, at the perturbative order we are considering, $[\vnabphi\cdot \dvf]$ does not mix with $\bigl(\Othree\bigr)\cdot \dvf$.

We have therefore found that the next-to-leading order renormalization correction reads

\begin{eqnarray}
\label{bbNLO}
\left[\vec{\nabla}\Phi\rt\cdot \dvfrvt\right]^{\rm NLO}&=&\left( \nu\sigma_0  (\Lambda)\left[\frac{1}{2 \sigma_0^2(\Lambda)}+ \right.\right.\nonumber\\
&+& \left.\left.\frac{31 \sigma_0^4(\Lambda)}{6 \left[-17\sigma_0^6(\Lambda) + 7 \mathcal K_1(\Lambda) \sigma_1^2(\Lambda)\right]}\right] \delta \vnabphi\right) \cdot \dvfrvt.\nonumber\\
\end{eqnarray}

\vspace{3mm}

Let us summarize the main results of this section, as they are the central point of this work: we have renormalized the composite operator $[\vnabphi\cdot \dvf]$ imposing the peak constraint. Already at first order, we have found that while the mean of $\vnabphi$ is zero as in the unconstrained case, the variance has an additional term. The composite operator therefore renormalized in such a way as to generate the correct variance and account for this statistical effect, in \eqref{bb}. We then compute the next-to-leading order effect and find the corresponding renormalization, \eqref{bbNLO}.

\subsection{Non-renormalization of the DM Boltzmann equation}
\label{sec:renorm ho}

\hspace{0.8cm} 
It is crucial to notice that for the smooth component of DM (that is without peak constraint) no renormalization is needed and the corresponding Boltzmann equation is not altered. This is consistent with the fact that we have used standard perturbation theory results for DM to evaluate the statistical correlators at the peak locations. 
Consistency can be explicitly checked upon noticing that, for DM, probabilities are not conditional. For  instance, we get at linear order
\begin{eqnarray}
\Big<\vec{\nabla}\Phi\rt\, f\rpt\Big>&=& f\rpt\,
\Big<\vec{\nabla}\Phi\rt \Big>=0, \nonumber\\
\Big<\vec{\nabla}\Phi\rt \, f(\vec{r},\vec{v}_1,t)\cdot \vec{\nabla}\Phi\rt \, f\rpvt\Big>&=&\Big<f(\vec{r},\vec{v}_1,t) \, f\rpvt\Big>\Big <\vec{\nabla}\Phi\rt \cdot \vec{\nabla}\Phi\rt\Big>\nonumber\\
&=&\frac{\alpha^2}{3}\sigma^2_{-1} \Big<f(\vec{r},\vec{v}_1,t)\,  f\rpvt\Big>
\end{eqnarray}
and one simply finds 
\be
\left[\vec{\nabla}\Phi\rt\right]^{\rm linear}_{\rm }=\vec{\nabla}\Phi\rt.
\ee
This remains true also at higher orders.

\section{The halo Boltzmann equation: the right-hand side}
\label{sec:boltzmann eq RHS}
We will now deal with the right-hand side of the Boltzmann equation, see Eqs.~(\ref{force}) and (\ref{irr}). 
It contains a force term that depends on the irreducible two-particle correlation (which, in turns, depends on the irreducible three-particle correlation etc.).
It is the correlation  encoded by $f_{\rm 2c}$ that causes the non-conservation of the one-particle phase-space distribution function in  configuration space. The physical interpretation of this effect is that the gravitational interactions between particles  (which induce long-range correlations) lead to clustering and, therefore, non-conservation of the one-particle phase-space distribution. 

To calculate this extra  force, we follow Ref.~\cite{bema}. The procedure consists in relating  phase-space densities to probability distributions for the mass and velocity field, which are naturally specified by models of cosmological structure formation. 
%
As explained in sec.~\ref{sec:boltzmann equation}, we can express the phase space density $f(\vr,\vv,t)$  in terms of the probability distribution $P(\rho,\vv,t)$. 
This approach can be extended to the irreducible two-particle correlation. Inspecting   eqs.~\eqref{force} and \eqref{irr}, we observe that
$f_{\rm 2c}(\vr,\vp,\vrp,\vvp,t)$ is integrated over $\d^3 v'$ and we needs only the velocity at one point (the superscript $'$ indicates that quantities are evaluated at $\vrp$)
\begin{eqnarray}
\label{as}
f_{\rm 2c}(\vr,\vp,\vrp,t)&=&\int\d^3 v'\,f_{\rm 2c}(\vr,\vp,\vrp,\vvp,t) \nonumber \\
&=&\int_0^\infty\,\d\rho\,\rho\, 
\int_0^\infty\,\d\rho'\,\rho'\,\left[ P(\rho,\vp,\rho',t)-P(\rho,\vp,t)P(\rho',t)\right]\nonumber\\
&=& \int_0^\infty\,\d\rho\,\rho\, 
\int_0^\infty\,\d\rho'\,\rho'\,\left[ P(\rho,\rho'|\vp,t)-P(\rho|\vp,t)P(\rho',t)\right]P(\vp,t)\nonumber\\
&=&\left[
\langle\rho\rho'|\vp\rangle-\langle\rho|\vp\rangle\langle\rho'\rangle
\right]P(\vvp,t)\;,
\end{eqnarray}
where, in the last equality, the averages are equal-time correlators.
Even though one derives these equations assuming a
single-valued velocity field in each realization, they are also
valid when there is a distribution of velocities at each point. We
simply interpret $\rho$  as the total mass density while $\vp$  is a single bulk 
velocity. Therefore, these equations are fully general \cite{bema}.

Let us specialize to the case of halos. The phase-space distribution (\ref{f integral rho}) can be written using Bayes' theorem as 
\be
\label{fpk}
f_{\rm pk}(\vr,\vv,t)=\int_0^\infty\,\d\rho\,\rho\, P(\vp|\rho,t)\,P(\rho,t)=\rho_{\rm pk}\, P(\vv,t | {\rm pk})=
				 \bar{\rho}(1+\delta_{\rm pk})\; J_{\vv\vpsi}^{-3}\,P(\vpsi,t | {\rm pk})
\ee
where, in the last passage, we have introduced the quantity
\be
\vec{\psi}\rt=-\frac{1}{\alpha}\vec{\nabla}\Phi\rt=-\int\frac{\d^3 r'}{4\pi}\, \frac{(\vr-\vrp)}{|\vr-\vrp|^3}\delta_{\rm }(\vrp,t),
\ee
and $J_{\vv\vpsi}$ is the Jacobian describing the passage from the variable $\vv$ to the variable $\vec{\psi}$.

In a similar fashion, we can compute the irreducible two-point correlation function
\begin{eqnarray}
\label{ad}
f^{\rm pk}_{\rm 2c}(\vr,\vp,\vrp,t)&=& \int_0^\infty\,\d\rho\,\rho\, 
\int_0^\infty\,\d\rho'\,\rho'\,\left[ P(\vv,\rho'|\rho,t)P(\rho,t)-P(\vv|\rho,t)P(\rho,t)P(\rho',t)\right]\nonumber\\
&=&\rho_{\rm pk}\int_0^\infty\,\d\rho'\,\rho'\, \left[P(\vv,\rho',t|{\rm pk})-P(\vv,t|{\rm pk})P(\rho',t)
\right]\nonumber\\
&=&\rho_{\rm pk}\, P(\vv,t|{\rm pk})\,\int_0^\infty\,\d\rho'\,\rho'\, \left[P(\rho',t|\vv,{\rm pk})-P(\rho',t)
\right]\nonumber\\
&=&\bar{\rho}^2(1+\delta_{\rm pk})\; J_{\vv\vpsi}^{-3}\,P(\vpsi,t | {\rm pk})\left[\Big<\delta(\vrp,t)\Big|\vpsi\rt,{\rm pk}\Big>-\Big<\delta(\vrp,t)\Big>\right]
\nonumber\\
&=&\bar\rho\; f_{\rm pk}(\vr,\vv,t)\,\Big<\delta(\vrp,t)\Big|\vpsi\rt,{\rm pk}\Big>\;.
\end{eqnarray}
In the last equality, we have substituted $\big<\delta(\vrp,t)\big>=0$ as this average vanishes in the absence of peak constraint.
The corresponding force felt by the halos is therefore
\be
\label{forcepeak}
\vec F_{\rm pk} = G_{\rm N}\bar\rho f_{\rm pk}(\vr,\vv,t)\,  \int \d^3 r'\,\Big<\delta(\vrp,t) \Big| \vpsi\rt,{\rm pk} \Big>\cdot \frac{\vr - \vrp}{|\vr - \vrp|^3}.
\ee
The computation of the force (\ref{forcepeak}) is reported in Appendix \ref{sec:appendix C} and we provide only the final result up to second-order
\begin{equation}
\label{eq:force peak result}
\begin{aligned}
\vec F_{\rm pk} (\vr,\vp,t) =& -4\pi\,G_{\rm N}\, \bar\rho\,f_{\rm pk}(\vr, \vp, t)\,\Biggl[\vec\psi\rt +\left(\vec\psi\rt -\frac{\sigma^2_0}{\sigma^2_1} \vec\nabla\delta\rt\right)\frac{\delta\rt}{\sss}\frac{\sigma_1^2}{\sigma_0^2}\\
	&\times \left(\Big<\delta\rt\vpsi\rt\cdot\vpsi\rt\Big>-\int\d \ell\,\frac{\ell_i}{\lvert\ell\lvert}\,\Big<\delta(\vrp,t)\delta\rt\psi_i\rt\Big> \right)\\
	&+\left(\vec\psi_1 -\frac{\sigma^2_{-1}}{\sigma^2_0} \vec\nabla\delta\rt\right)\frac{\delta\rt}{\sss} \biggl(\int\d \ell\,\frac{\ell_i}{\lvert\ell\lvert}\,\Big<\delta(\vrp,t)\delta\rt\partial_i\delta\rt\Big> \\
	&-\Big<\delta\rt\vpsi\rt\cdot\vec{\nabla}\delta\rt\Big> \biggr)\Biggr],
\end{aligned}
\end{equation}
where $\vec\ell=(\vrp-\vr)$. Using the results for the cumulants reported in \eqref{mu233}, \eqref{mu234}, \eqref{mu123}, \eqref{mu124} we find (notice that the last two lines in \eqref{eq:force peak result} cancel each other)
\begin{multline}
\label{eq:force peak result final}
\vec F_{\rm pk} (\vr,\vp,t) = -4\pi\,G_{\rm N}\, \bar\rho\,f_{\rm pk}(\vr, \vp, t)\,\Biggl[\vec\psi\rt
  +\left(\vec\psi\rt -\frac{\sigma^2_0}{\sigma^2_1} \vec\nabla\delta\rt\right)\\
  \times\frac{\delta\rt}{\sss}\frac{\sigma_1^2}{\sigma_0^2} \left(\frac 32 \mathcal K_1 -\frac{10}{21}\sigma_0^2 \sigma_{-1}^2\right)\Biggr].
\end{multline}
Two comments are in order. Firstly, we can easily recover the DM case away from the peaks, either by computing directly the force from expression (\ref{ad}) with aid of $\langle\delta(\vrp,t)\vpsi\rt\rangle$, or by eliminating the peak constraint taking $\sigma_0^2,\sigma_1^2\gg 1$. In both cases, we obtain
\begin{equation}
\vec F_{\rm dm} (\vr,\vp,t) = -4\pi\,G_{\rm N}\, \bar\rho\,f_{\rm dm}(\vr, \vp, t)\,\vec\psi\rt.
\end{equation}
Secondly, neither the force on the peaks nor on the DM contributes to the momentum equation once the integration of the force over momenta is considered,
\be
\int\d^3 v\, \vv\, \vec\nabla_\vv\cdot\vec F_{\rm pk,dm} (\vr,\vp,t) =\vec 0,
\ee
in agreement with the Equivalence Principle. Technically, this cancellation follows from the fact that integrating out the velocities is equivalent to an integration over the configurations of the field $\vpsi$. As the phase-space distribution is proportional to $P(\vpsi,t)$, we obtain
\be
\int\d^3\psi\,P(\vpsi,t)\, \vpsi=\big<\vpsi\big>=\left\{
\begin{aligned}
& -\alpha\,\frac{\sigma_0^2(\Lambda)}{\sigma_1^2(\Lambda)}\,\left.\vec{\nabla}\delta_{\rm }\right|_{\rm pk}=\vec 0 && (\text{for peaks})\,, \\
& \ \vec 0 			& & 			 (\text{for DM})\,,
\end{aligned}\right.
\ee
where, for peaks, we made use of eq.~(\ref{average grad phi}). 
This equation seems to suggest that one would break the equivalence principle if the constraint were different from $\vec{\nabla}\delta=\vec 0$, as may be the case for other tracers of the large scale structure. However, isotropy and homogeneity would enforce a constraint on the modulus of $\vec{\nabla}\delta$ only, e.g. $|\vec{\nabla}\delta|\leq c$ for some constant $c>0$. Therefore, the integral over $\vpsi$ would also vanish in those cases, and the Equivalence Principle would still be satisfied, as it should be.

\section{Discussion and conclusion}
\label{sec:conclusions}
The fine understanding of the large scale structure of our Universe is a central theme in cosmology. 
Our knowledge in the field is nowadays (and will be even more in the future) dominated by data.  
It is, therefore, essential to understand in detail the statistical properties of the observed galaxies. 
Achieving this goal automatically calls for a detailed investigation of the statistical properties of the 
DM halos, where galaxies are expected to reside.

In this paper we have derived an effective Boltzmann equation that describes the dynamics of the halo 
mean-field phase space. 
This equation is necessarily  different from the Boltzmann equation of DM as halos are not statistically 
unbiased relatively to the coarse-grained DM density field. 
Through a renormalization procedure of composite operators, we have obtained up to second-order in 
perturbation theory the statistically biased gravitational force felt by halos. 
We conclude that, at this order, the Boltzmann Equation for halos reads
\begin{multline}\label{final}
\frac{\partial f_{\rm pk}}{\partial t}+\vec{v}\cdot \frac{\partial f_{\rm pk}}{\partial\vr} 
-\Biggl(\vnabphi\ww\rt+\alpha\frac{\sigma_0^2(R)}{\sigma_1^2(R)}\vec{\nabla}\delta\ww\rt + \\
+ \nu\sigma_0(R)\left[\frac{1}{2 \sigma_0^2(R)} + \frac{31 \sigma_0^4(R)}{6 \left[-17\sigma_0^6(R) + 7 \mathcal K_1(R) \sigma_1^2(R)\right]}\right] \delta\ww\rt \vnabphi\ww\rt\Biggr)\cdot\frac{\partial f_{\rm pk}}{\partial\vp}=0 \;,
\end{multline}
where the extra contributions to $\vnabphi\ww\rt$ are purely of statistical origin: they do not 
arise because halos feel a different gravitational force than dark matter, as this would violate the equivalence principle, but rather as a result of imposing the peak constraint on their statistics\footnote{ An alternative, perhaps more intuitive, interpretation of this result is the following: instead of modifying the Boltzmann equation for halos with an effective force field, one may interpret the results of equations \eqref{bb} and \eqref{bbNLO} as the corrections to the probability distribution function (PDF) of the force acting on halos. In this picture, the Boltzmann equation for halos \eqref{final} would look exactly as the one of DM, equation \eqref{eq:dmfinal}, but with a conditional PDF of the form of the terms in bracket in equation \eqref{final}. }.
This should be contrasted to the DM Boltzmann equation
\be\label{eq:dmfinal}
\frac{\partial f_{\rm }}{\partial t}+\vec{v}\cdot \frac{\partial f_{\rm }}{\partial\vr}
-\vec{\nabla}\Phi\rt\cdot\frac{\partial f_{\rm }}{\partial\vp}=0 \;.
\ee
We emphasize again that $\fpk$ is the one-particle phase space density associated with the mean-field 
distribution of halos that corresponds to a particular $f$, that is, a particular realization of the large 
scale structure. 
The biased gravitational force experienced by the halo centers imprints a signature, among others, in the 
peculiar velocities of virialized halos. 
This generates the first-order statistical velocity bias which has been discussed in \cite{v0,dcss,v1} and 
measured in \cite{porciani,v1}.
However, tidal forces etc. will also be affected. Our approach can in principle be applied to work out these
corrections at any order in perturbation theory.

\begin{figure}[h!]
\centering
\includegraphics[width=0.8\textwidth]{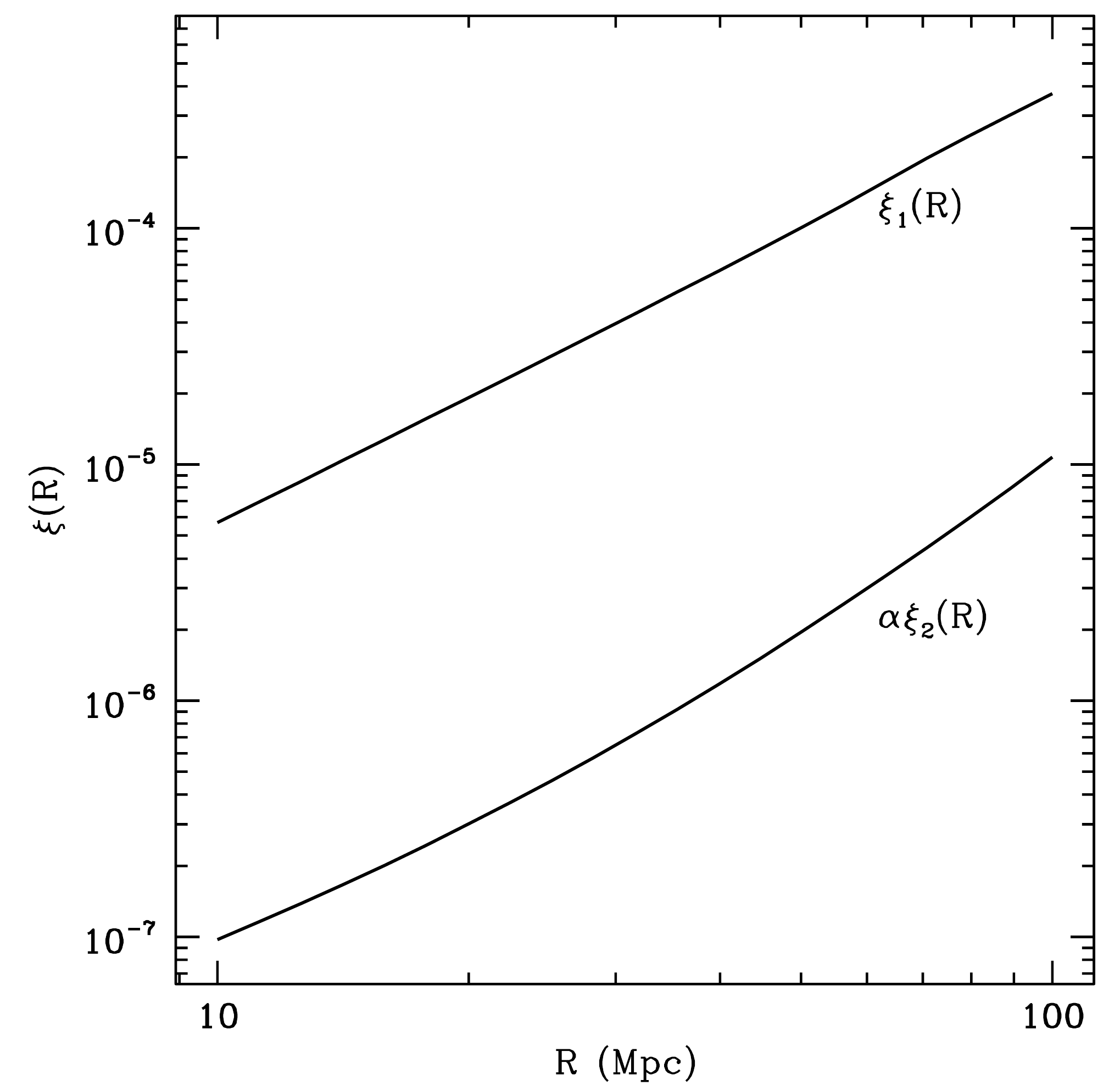}
\caption{$\xi_1(R)$ and $\alpha\xi_2(R)$ plotted here as a function of the Lagrangian scale $R = 1/ \Lambda$. Here we normalize $\xi_2(R)$ with a factor of $\alpha$ which relates $\delta$ with $\nabla \phi$ in \eqref{final}.}
\label{xir}
\end{figure}

To give intuition about the magnitude of the statistical correction to the gravitational force in 
eq.~\eqref{final}, we plot in fig. \ref{xir} the amplitude of the second and third terms in parenthesis
\begin{eqnarray}
\xi_1(R) &=& \alpha\frac{\sigma_0^2(R)}{\sigma_1^2(R)},\\
\xi_2(R) &=& \nu\sigma_0(R)\left[\frac{1}{2 \sigma_0^2(R)} + \frac{31 \sigma_0^4(R)}{6 \left[-17\sigma_0^6(R) + 7 \mathcal K_1(R) \sigma_1^2(R)\right]}\right]
\end{eqnarray}
as a function of the Lagrangian scale $R = 1/\Lambda$ assuming the biasing is described by the peak 
constraint.
The magnitude of the second-order statistical correction also rises with $R$, in agreement with the 
expectation that these effects should increase with halo mass.

Measurements from N-body simulations are challenging since the effect must vanish in the limit $k\to 0$. 
Furthermore, the sparsity of massive halos could bias the measurements when the purpose is to define 
volume-weighted statistics (see e.g. \cite{zzj2} for recent discussions). 
Note, however, that number-weighted statistics are most suited to extract these effects as they are the
quantities being observed.

There is some confusion in the literature regarding the nature of this effect. A physical bias will immediately 
arise as soon as there are at least two different fluids since, in this case, velocities do not need to share
the same value. This bias is effective on an object-by-object basis.
This is obviously the case of e.g. baryons and DM. Furthermore, this will also be the case of 
the halos and DM since, owing to the finite halo size, the force acting on the halo center-of-mass, $\vnabphi\ww$, 
is different from the force acting on DM particles, $\vnabphi$. 
Such a coarse-graining procedure can introduce additional terms in the fluid equations, as was already pointed 
out in \cite{buchert}. This possibility was recently reconsidered in \cite{chan1}. 
If we ignore in Eq.(\ref{final}) all the statistical corrections but the term $\vnabphi\ww$, then our results 
are identical to his, except for the time constancy of our $R$. 
As shown in \cite{chan1}, a time-dependent filter $W(\vr/R,t)$ would yield extra terms in the moments of the 
Boltzmann equation owing to the time derivative $\partial f/\partial t$.

However, \cite{chan1} did not include the statistical corrections, which are the focus of this work. In fact,
our results would hold even in the extreme case where no extra filtering is applied to the macrostate. This 
illustrates the fact that the statistical effects we are discussing are {\it not} caused by the finite extent 
of the halos.
Regarding the existence of these statistical effects, we note that, while a time-dependent window $W(\vr/R,t)$ 
mimicking the collapse of halos can induce $k$-dependent contributions at low redshift, corrections are small 
at high redshift \cite{chan1}. Therefore, this cannot be the explanation for the suppression of the proto-halo 
velocity power spectrum measured in \cite{porciani,v1}.

To conclude, note that halo number conservation implies that we follow the collapse of Lagrangian ``peak-patch'', 
as coined by \cite{bond/myers:1996}, into parent or isolated DM halos. Therefore, we do not explicitly track the 
merging history of these host halos. Although one could naively apply the extended Press-Schechter (EPS) approach
\cite{bond/cole/etal:1991} to determine the merger history, it would be very interesting to assess whether the 
statistical effects considered here also affect merger rates etc. However, it is not obvious how to do this within 
our current formulation, especially since the smoothing scale $R$ is linked to the parent halo mass and, thus, is 
not free unlike in EPS theory. Our assumption that the density peak retain their initial critical height (i.e. 
$\nu$ does not depend on time) is another limitation of our approach. A better treatment would consistently 
track the time evolution of both $R$ and $\nu$ such that the mass enclosed within $R$ always equates the host halo
mass. We leave all this for future work.

%
\acknowledgments
We acknowledge an anonymous referee for her/his careful reading and helpful suggestions.
The research of A.K. was implemented under the Aristeia Action of the Operational Programme Education and 
Lifelong Learning and is co-funded by the European Social Fund (ESF) and National Resources. 
A.R. is supported by the Swiss National Science Foundation (SNSF), project `The non-Gaussian Universe" 
(project number: 200021140236). 
M.B. and V.D. acknowledge support by the SNSF.

%
\appendix

\section{Conditional probabilities}
\label{sec:appendix A}
\hskip 0.8cm In this Appendix we present the techniques we use to compute conditional probabilities of initially Gaussian fields   which develope non-Gaussian fluctuations as time evolves. This method is inspired by the path integral approach and a similar application can be found in \cite{MR3}.

Let us suppose there are $n$ stochastic variables $X_i$, with $i=1,\cdots,n$. We define the connected two-point correlators as
\be
\mu_{ij}=\langle X_i X_j\rangle_{\rm c}
\ee
and the connected three-point correlators as
\be
\mu_{ijk}=\langle X_i X_j X_k\rangle_{\rm c}.
\ee
Thanks to Bayes' theorem, we can define the conditional probability that $X_1$ assumes a value given that the other variables assume defined values as
\be
P(X_1|X_2,\dots,X_n)=\frac{P(X_1, X_2,\dots,X_n)}{P(X_2,\dots,X_n)},
\ee
where
\be
P(X_2,\dots,X_n)=\int\d X_1\,P(X_1, X_2,\dots,X_n).
\ee
The basic quantity to compute is therefore $P(X_1, X_2,\cdots,X_n)$. Following Ref. \cite{MR3}, it can be expressed in the following way
\be
P(X_1, X_2,\dots,X_n)=\int{\cal D}\lambda\, e^{i\lambda_i X_i}\,e^{-\frac{1}{2}\lambda_i\lambda_j  \mu_{ij}}\,e^{\frac{(-i)^3}{6}\lambda_i\lambda_j\lambda_k  \mu_{ijk}} \cdots\, ,
\ee
where
\be
{\cal D}\lambda\equiv \int_{-\infty}^\infty\frac{\d \lambda_1}{2\pi}\cdots \int_{-\infty}^\infty\frac{\d \lambda_n}{2\pi}.
\ee
Using the fact that 
\be
\lambda e^{i\lambda X}=-i\p_X e^{i\lambda X},
\ee
perturbing for small three-point correlators and defining with subscript ${}_{\rm g}$ the Gaussian counterparts, we find 
\begin{equation}
\label{gen}
\begin{aligned}
P(X_1|X_2,\dots,X_n)=\,&P_{\rm g}(X_1|X_2,\dots,X_n)\\
&-\frac{\mu_{ijk}}{6P_{\rm g}(X_2,\dots,X_n)}\,\p_i\p_j\p_k\,P_{\rm g}(X_1, X_2,\dots,X_n)\\
&+\frac{P_{\rm g}(X_1, X_2,\dots,X_n)}{6P^2_{\rm g}(X_2,\dots,X_n)}
\mu_{ijk}\int\d X_1\,\p_i\p_j\p_k\,P_{\rm g}(X_1, X_2,\dots,X_n).
\end{aligned}
\end{equation}
The Gaussian probability can be expressed as 
\be
P_{\rm g}(X_1, X_2,\dots,X_n)=\frac{1}{(2\pi)^{n/2}}\frac{1}{\left({\rm Det}\, C\right)^{1/2}}\,e^{-\frac{1}{2} X_i C^{-1}_{ij} X_j},
\ee
where $C$ is the covariance matrix. 
We now write
\begin{equation}
\begin{aligned}
\mu_{mqr}\p_m\p_q\p_r =&\ \mu_{111}\p_1^3\\
&+3\hat\mu_{1qr}\p_1\p_q\p_r \\
&+3\hat\mu_{11r}\p^2_1\p_r \\
&+\hat\mu_{mqr}\p_m\p_q\p_r \,,
\end{aligned}
\end{equation}
where the hat in the correlator $\hat\mu_{mqr}$ means that we are taking indices different from $1$.\\
Using equation \eqref{gen} we get
\begin{equation}
\label{x1 general}
\begin{aligned}
\langle X_1| X_2,\dots,X_n\rangle =&\int\d X_1\,X_1\,P(X_1|X_2,\dots,X_n) \\
=&\ \langle X_1| X_2,\dots,X_n\rangle_\text{g} \\
&- \frac{\mu_{111}}{6P_{\rm g}(X_2,\dots,X_n)}\,\int\d X_1\,X_1\, \p_1^3P_{\rm g}(X_1, X_2,\dots,X_n) \\
&-\frac{3\hat\mu_{1qr}}{6P_{\rm g}(X_2,\dots,X_n)}\,\int\d X_1\,X_1\, \p_1\p_q\p_rP_{\rm g}(X_1, X_2,\dots,X_n) \\
&-\frac{3\hat\mu_{11r}}{6P_{\rm g}(X_2,\dots,X_n)}\,\int\d X_1\,X_1\, \p^2_1\p_rP_{\rm g}(X_1, X_2,\dots,X_n) \\
&-\frac{\hat\mu_{mqr}}{6P_{\rm g}(X_2,\dots,X_n)}\,\int\d X_1\,X_1\, \p_m\p_q\p_rP_{\rm g}(X_1, X_2,\dots,X_n) \\
&+\frac{\langle X_1| X_2,\dots,X_n\rangle_\text{g}}{6P_{\rm g}(X_2,\dots,X_n)}
\mu_{111}\int\d X_1\,\p^3_1\,P_{\rm g}(X_1, X_2,\dots,X_n) \\
&+\frac{\langle X_1| X_2,\dots,X_n\rangle_\text{g}}{6P_{\rm g}(X_2,\dots,X_n)}
3\hat\mu_{1qr}\int\d X_1\,\p_1\p_q\p_r\,P_{\rm g}(X_1, X_2,\dots,X_n) \\
&+\frac{\langle X_1| X_2,\dots,X_n\rangle_\text{g}}{6P_{\rm g}(X_2,\dots,X_n)}
3\hat\mu_{11r}\int\d X_1\,\p_1^2\p_r\,P_{\rm g}(X_1, X_2,\dots,X_n) \\
&+\frac{\langle X_1| X_2,\dots,X_n\rangle_\text{g}}{6P_{\rm g}(X_2,\dots,X_n)}
\hat\mu_{mqr}\int\d X_1\,\p_m\p_q\p_r\,P_{\rm g}(X_1, X_2,\dots,X_n),
\end{aligned}
\end{equation}
that can be simplified in
\begin{equation}
\label{x1}
\begin{aligned}
\langle X_1| X_2,\dots,X_n\rangle=&\int\d X_1\,X_1\,P(X_1|X_2,\dots,X_n) \\
=&\ \langle X_1| X_2,\dots,X_n\rangle_\text{g} \\
&+ 0 \\
&+ \frac{\hat\mu_{1qr}}{2P_{\rm g}(X_2,\dots,X_n)}\p_q\p_r P_{\rm g}(X_2,\dots,X_n) \\
&+ 0 \\
&- \frac{\hat\mu_{mqr}}{6P_{\rm g}(X_2,\dots,X_n)}\p_m\p_q\p_r \int\d X_1\,X_1\, P_{\rm g}(X_1, X_2,\dots,X_n) \\
&+ 0 \\
&+ 0 \\
&+ 0 \\
&+ \frac{\langle X_1| X_2,\dots,X_n\rangle_\text{g}}{6P_{\rm g}(X_2,\dots,X_n)}
\hat\mu_{mqr}\p_m\p_q\p_r P_{\rm g}(X_2,\dots,X_n).
\end{aligned}
\end{equation}
Similarly,
\begin{equation}
\label{x1sq}
\begin{aligned}
\langle X_1^2| X_2,\dots,X_n\rangle=&  \int\d X_1\,X^2_1\,P(X_1|X_2,\dots,X_n) \\
=&\   \langle X_1^2| X_2,\dots,X_n\rangle_\text{g} \\
&+ \frac{\hat\mu_{1qr}}{P_{\rm g}(X_2,\dots,X_n)}\p_q\p_r \int\d X_1\,X_1\, P_{\rm g}(X_1, X_2,\dots,X_n) \\
&- \frac{\hat\mu_{11r}}{P_{\rm g}(X_2,\dots,X_n)}\p_r  P_{\rm g}( X_2,\dots,X_n) \\
&- \frac{\hat\mu_{mqr}}{6P_{\rm g}(X_2,\dots,X_n)}\p_m\p_q\p_r \int\d X_1\,X^2_1\, P_{\rm g}(X_1, X_2,\dots,X_n) \\
&+ \frac{\langle X_1^2| X_2,\dots,X_n\rangle_\text{g}}{6P_{\rm g}(X_2,\dots,X_n)}
\hat\mu_{mqr}\p_m\p_q\p_r P_{\rm g}(X_2,\dots,X_n).
\end{aligned}
\end{equation}

\section{Cumulants at second-order}
\label{sec:appendix B}
\hskip 0.8cm In Appendix \ref{sec:appendix A} we have showed how to calculate the average of a stochastic variable which is conditioned by a number of other variables, taking into account its non-Gaussian nature. 

In this Appendix we apply such a  technique to compute the average of $\vec\nabla\Phi$ on the peak up to second-order. This amounts to using \eqref{x1}, which requires the knowledge of the the cumulants $\mu_{mqr}=\langle X_m\, X_q\,X_r\rangle$,
where the $X_i$ range among the seven Gaussian variables $\delta$, $\vec\nabla_i\Phi$, and  $\vec\nabla_i\delta$ ($i=1,2,3$). 
Here we are omitting the dependence on $r$ and $t$ as the variables are all evaluated at the same point. 
Because of rotational invariance, the only non vanishing cumulants are
\begin{gather}
\langle \delta\rt\,\delta\rt\,\delta\rt \rangle, \qquad \langle \delta\rt\,\nabla_i\Phi\rt\,\nabla_j\Phi \rt\rangle,\nonumber\\
\langle \delta\rt\,\nabla_i\Phi\rt\,\nabla_j\delta\rt \rangle, \qquad
\langle \delta\rt\,\nabla_i\delta\rt\,\nabla_j\delta\rt \rangle.
\end{gather}
We show the explicit computation of one of them, the others are calculated in a similar fashion. Let us take for example $\langle \delta\nabla_i\Phi\,\nabla_j\delta \rangle$. Going to Fourier space we have
\begin{equation}
\begin{aligned}
\langle \delta\,\nabla_i\Phi\,\nabla_j\delta\rangle =& \int \frac{\d^3 k_1}{(2\pi)^3}\frac{\d^3 k_2}{(2\pi)^3}\frac{\d^3 k_3}{(2\pi)^3}\frac{\d^3 q}{(2\pi)^3}\Bigl[ \langle\delta(\vq)\,\delta(\vk_1-\vq)\,\delta(\vk_2)\,\delta(\vk_3)\rangle F_2(\vq,\vk_1-\vq)+ \\
& +\text{cyc.}\Bigl]\, \left(i\,\alpha\, \frac{k_{2_i}}{k^2_2}\right)\,\left(-i k_{3_j}\right) e^{-i (\vk_1+\vk_2+\vk_3)\cdot \vr} \\
=& \ \alpha\,\int \frac{\d^3 k_2}{(2\pi)^3}\frac{\d^3 k_3}{(2\pi)^3} \,P(k_2)\,P(k_3)\,2F_2(\vk_2,\vk_3)\,\left( \frac{k_{2_i}k_{3_j}}{k^2_2}\right)\, \\
&- \alpha  \int \frac{\d^3 k_1}{(2\pi)^3}\frac{\d^3 k_3}{(2\pi)^3} \,P(k_1)\,P(k_3)\,2F_2(\vk_1,\vk_3)\,\left( \frac{(k_{1_i}+k_{3_i})k_{3_j}}{|\vk_1 + \vk_3|^2}\right) +  \\
&- \alpha \int \frac{\d^3 k_1}{(2\pi)^3}\frac{\d^3 k_2}{(2\pi)^3} \,P(k_1)\,P(k_2)\,2F_2(\vk_1,\vk_2)\,\left( \frac{k_{2_i} (k_{1_j}+k_{2_j})}{k^2_2}\right), 
\end{aligned}
\end{equation}
where we have used the usual standard perturbation theory kernel
\begin{equation}
F_2(\vq_1, \vq_2) = \frac 57 + \frac 12 \frac{\vq_1\cdot \vq_2}{q_1 q_2}\left(\frac{q_1}{q_2}+\frac{q_2}{q_1}\right) + \frac 27 \left(\frac{\vq_1\cdot \vq_2}{q_1 q_2}\right)^2
\end{equation}
and the Poisson equation 
\be
\nabla^2 \Phi = \alpha \delta.
\ee
Let us define $A$ by
\begin{equation}\label{adel}
\langle \delta\,\nabla_i\Phi\,\nabla_j\delta\rangle = A\, \delta_{ij}\,.
\end{equation}
By taking the trace we get
\begin{equation}
A=-\frac {\alpha}{3} \int \frac{\d k_1}{2\pi^2}\frac{\d k_2}{(2\pi)^2} k^2_1\,k^2_2\,P(k_1)\,P(k_2)\, \int \d \mu \,2F_2(\vk_1,\vk_2)\,\left[ 1+\frac{(\vk_1+\vk_2)\cdot \vk_2}{|\vk_1+\vk_2|^2}\right],
\end{equation}
where $\mu=\cos\theta$, being $\theta$ the angle between $\vk_1$ and $\vk_2$. We can now integrate over $\mu$ and obtain
\begin{equation}
A=-\alpha\,\frac{17}{21}  \int \frac{\d k_1}{2\pi^2}\frac{\d k_2}{2\pi^2} k^2_1\,k^2_2\,P(k_1)\,P(k_2) = -\alpha\,\frac{17}{21} \sigma^4_0.
\end{equation}
A similar calculation of the other cumulants gives
\begin{eqnarray}
\left\langle \delta\,\delta\,\delta \right\rangle &=& \frac{34}{7} \sigma^4_0 \,,\label{mu222}\\
\left \langle \delta\,\nabla_i\Phi\,\nabla_j\Phi \right\rangle &=&\frac{\alpha^2}{3}\, \mathcal K_1 \, \delta_{ij}\,, \label{mu233}\\
\left\langle \delta\,\nabla_i\Phi\,\nabla_j\delta \right\rangle  &=&-\alpha\, \frac{17}{21} \sigma^4_0 \, \delta_{ij} \,, \label{mu234}\\
\left\langle \nabla_i \Phi\, \nabla_l\Phi \, \nabla_l\nabla_j\Phi \right\rangle  &=& -\frac{\alpha}{2}  \left \langle \delta\,\nabla_i\Phi\,\nabla_j\Phi \right\rangle =-\frac{\alpha^3}{6}\, \mathcal K_1 \, \delta_{ij} \,, \label{corr 03}  \\
\left\langle \nabla_i \delta\, \nabla_l\Phi \, \nabla_l\nabla_j\Phi \right\rangle  &=&  - \frac{\alpha^2}{3}\, \mathcal K_2 \, \delta_{ij} \,. \label{corr 13}
\end{eqnarray}
Here we have defined the  quantities
\begin{small}
\begin{equation}
\label{K1,K2}
\begin{aligned}
\mathcal K_1=\int & \frac{\mathrm dk_1\mathrm dk_2}{4\pi^4} P(k_1)P(k_2) \times \\
	\times&\left[ \frac{5\left[ 2k_1k_2\Bigl(k_1^2+k_2^2\Bigr)\Bigl(3k_1^4-14 k_1^2k_2^2+3k_2^4\Bigr) +3 (k_1^2-k_2^2)^4 \left(\log \frac{|k_1-k_2|}{k_1+k_2}\right)\right]}{168 \, k_1^3 k_2^3} \right] \,, \\
\mathcal K_2 =
\int & \frac{\mathrm dk_1\mathrm dk_2}{4\pi^4} P(k_1)P(k_2) \Biggl[ \frac 75 k_1^2 k_2^2  + \\ 
+ & \Bigl(k_1^2+k_2^2\Bigr) \frac{5\left[ 2k_1k_2 \Bigl(k_1^2+k_2^2\Bigr)\Bigl(3k_1^4-14 k_1^2k_2^2+3k_2^4\Bigr) +3 (k_1^2-k_2^2)^4 \left(\log \frac{|k_1-k_2|}{k_1+k_2}\right)\right]}{336 \, k_1^3 k_2^3} \Biggr]\,.
\end{aligned}
\end{equation}
\end{small}
\noindent
These quantities are finite: the integrand is finite for $k_1\to k_2$, and for $k_1$ or $k_2\to 0$ the power spectra lead to a finite result.

We are now ready to compute the average of $\vec\nabla\Phi$ up to second-order. We use \eqref{x1} which contains the Gaussian probabilities, encoded by the covariance matrix for the variables $(\delta, \vec \nabla \Phi, \vec \nabla \delta)$ (we understand the symmetric components of the covariance matrix)
\begin{equation}
\label{covariance matrix 234}
\mathcal C=
\begin{pmatrix}
 \sigma_0^2	&	& \vec 0						& & \vec 0\\\\
 						& & \dfrac{\alpha^2}{3} \sigma_{-1}^2 \mathbb 1_3 & & -\dfrac{\alpha}{3} \sigma_{0}^2 \mathbb 1_3 \\\\
 	&&  && \dfrac 13 \sigma_{1}^2 \mathbb 1_3
\end{pmatrix}.
\end{equation}
By using \eqref{adel} for all the cumulants we find
\begin{equation}
\begin{aligned}
\langle \vec\nabla\Phi\rangle_{\rm pk} = & -\alpha\,\frac{\sigma^2_0}{\sigma^2_1} \vec \nabla\delta 
     +\frac{\delta\vec \nabla\delta}{\sigma^2_0\sigma^2_1}\left(\langle \delta\,\vec\nabla\Phi\cdot\vec \nabla\delta \rangle 
     +\alpha\langle \delta\,\vec\nabla\delta\cdot\vec\nabla\delta \rangle\frac{\sigma^2_0}{\sigma^2_1}\right)\\
	=& -\alpha\,\frac{\sigma^2_0}{\sigma^2_1} \vec\nabla\delta +\alpha\,\frac{31}{21} \frac{\sigma^2_0}{\sigma^2_1}\delta\vec\nabla\delta.
\end{aligned}
\end{equation}
In a similar fashion, we can compute the covariance
\begin{equation}
\begin{aligned}
\langle \nabla_i\Phi\cdot\nabla_j\Phi\rangle_{\rm pk}  =& \frac 13\delta_{ij} \Biggl[\alpha^2\left(\sigma^2_{-1} - \frac{\sigma^4_0}{\sigma^2_1}\right) \\ 
	+& \,\langle \delta\,\vec\nabla\Phi\cdot\vec\nabla\Phi \rangle \, \frac{\delta_{\rm pk}}{\sigma^2_0} 
	+\alpha^2 \langle \delta\,\vec\nabla\delta\cdot \vec\nabla\delta \rangle\, \delta_{\rm pk}\, \frac{\sigma^2_0}{\sigma^4_1} 
	+2\alpha\, \langle \delta\,\vec\nabla\Phi\cdot\vec\nabla\delta \rangle\, \frac{\delta_{\rm pk}}{\sigma^2_1}\Biggr] =\\
	=& \frac{\alpha^2}{3}\delta_{ij}\biggl[
	\left(\sigma^2_{-1} - \frac{\sigma^4_0}{\sigma^2_1}\right)+
	\delta_\text{pk}\left(\mathcal K_1\frac{1}{\sigma_0^2} -\frac{20}{21} \frac{\sigma_0^4}{\sigma_1^2}\right) \biggr].
\end{aligned}
\end{equation}

\section{The peak force}
\label{sec:appendix C}
In this section we report the intermediate results needed to compute the conditioned expected value $\langle\delta(\vrp ,t)\Big|\vpsi\rt,{\rm pk}\rangle$ appearing in \eqref{forcepeak}.

The formula needed to compute this conditioned probability is again \eqref{x1}. In this case, differently from what discussed in the previous section, also $\delta(\vrp)$ (or, shortly, $\delta'$) appears among the variables and hence in the covariance matrix. In order to write down the new covariance matrix corresponding to \eqref{covariance matrix 234} we order the variables as $\{\delta(\vrp),\delta(\vr), \vec \nabla \Phi(\vr), \vec \nabla \delta(\vr)\}$ and we denote $\vec \ell= (\vrp-\vr)$,
so that the covariance matrix reads 
\begin{equation}
\mathcal C=
\begin{pmatrix}
\sigma_0^2 && \xi(\ell)				&& \eta(\ell)\vl		&&  -\dfrac{\mathrm d\xi(\ell)}{\mathrm d \ell}\dfrac{\vl}{\ell} \\\\
					 && \sigma_0^2	&&	\vec 0						&& \vec 0\\\\
					 && 						&& \dfrac 13 \sigma_{-1}^2 \mathbb 1_3 && \dfrac 13 \sigma_{0}^2 \mathbb 1_3 \\\\
					 && 	&&  && \dfrac 13 \sigma_{1}^2 \mathbb 1_3
\end{pmatrix},
\end{equation}
where we have defined
\begin{gather}
\xi(r) = \int \frac{\d^3 k}{(2\pi)^3} P(k) j_0(kr)\,,\\
\eta(r) = \int \frac{\d^3 k}{(2\pi)^3} P(k) \frac{j_1(kr)}{kr}\,,
\end{gather}
and $j_n(x)$ are the spherical Bessel functions. The following formul\ae\ for their integrals over the solid angle will be needed:
\begin{gather}
\label{integrals xi eta}
\int \d^3 r\, \frac{r_i\,r_j}{r^3} \frac{1}{r}\frac{\d\xi}{\d r} = -\frac{4\pi}{3} \sigma^2_0 \delta_{ij},\\
\int \d^3 r\, \frac{r_i\,r_j}{r^3}\, \eta(r) =\frac{4\pi}{3} \sigma^2_{-1} \delta_{ij}.
\end{gather}
We can now write down the final result after the application of \eqref{x1}. We mark in blue the terms that vanish with the subsequent integration $\int\mathrm d^3\ell \,\vl$. We also denote the cumulants with a compact notation, so that for example $\muACD=\langle \delta(\vrp)\vec\psi(\vr) \cdot\vec\nabla \delta(\vr)\rangle$. 
\begin{footnotesize}
\begin{equation*}
\begin{aligned}
\Big<\delta(&\vrp, t)\Big| \vpsi\rt,\delta,\vec\nabla\delta\Big>
= \zero{\frac{\delta_1}{\sigma^2_0}\xi(\ell)}-\frac{3}{\sigma^4_0 - \sigma^2_1 \sigma^2_{-1}}\left( \sigma^2_1\,\eta(\ell) + \sigma^2_0\,\frac{1}{\ell}\,\frac{d\xi}{\d\ell} \,\right)\vl\cdot\vec\psi_1 \\
	\zero{+}& \zero{\frac{3}{\sss}\left(\sigma^2_{-1}\,\frac{1}{\ell}\,\frac{d\xi}{\d\ell} + \sigma^2_0\,\eta(\ell)\right)\, \vl\cdot\vec\nabla\delta_1 + \frac{1}{2}\,\muABB\, \frac{1}{\sigma^2_0}\left(\frac{\delta^2_1}{\sigma^2_0} -1\right)}\\
	+ & 3\,\muABC \cdot \Biggl[\vec\nabla\delta_1-\frac{\sigma^2_1 }{\sigma^2_0}\,\vec\psi_1\Biggl]\,\frac{\delta_1}{\sss} + 3\,\muABD\cdot \Biggl[\vec\psi_1-\frac{\sigma^2_{-1} }{\sigma^2_0}\,\vec\nabla\delta_1\Biggl]\,\frac{\delta_1}{\sss} \\	 
	\zero{+}& \zero{\frac{1}{2}\,\muACC \Biggl[\frac{3\,(\sigma^2_0\,\vec\nabla\delta_1-\sigma^2_1\,\vec\psi_1)^2}{(\sss)^2} +\frac{\sigma^2_1}{\sigma^4_0 - \sigma^2_1 \sigma^2_{-1}}\Biggl]+\frac{1}{2}\,\muADD \Biggl[\frac{3\,(\sigma^2_{-1}\,\vec\nabla\delta_1-\sigma^2_0\,\vec\psi_1)^2}{(\sss)^2}+\frac{\sigma^2_{-1}}{\sigma^4_0 - \sigma^2_1 \sigma^2_{-1}}\Biggl] }\\
	\zero{-}&\zero{\muACD\Biggl[ \frac{ 3\sigma^2_0\,(\sigma^2_1\,|\vec\psi_1|^2 + \sigma^2_{-1}\,|\vec\nabla\delta_1|^2)-3\vec\nabla\delta_1\cdot\vec\psi_1(\sigma^4_0+\sigma^2_1\sigma^2_{-1})}{(\sss)^2} +\frac{\sigma^2_0}{\sigma^4_0 - \sigma^2_1 \sigma^2_{-1}}\Biggl]}\\
	\zero{+}& \zero{\frac{1}{2}\,\muBBB\, \frac{1}{\sigma^4_0}\,\left(1-\frac{\delta^2_1}{\sigma^2_0}\right) \,\,\xi(\ell)} +\frac{1}{2}\,\muBCC\Biggl\{\frac{6\,\delta_1}{(\sss)^2}\,\left[\frac{1}{\ell}\frac{\d \xi(\ell)}{\d \ell}\sigma^2_0 + \eta(\ell)\sigma^2_1\right]\,\vl\cdot\left[\vec\nabla\delta_1-\frac{\sigma^2_1}{\sigma^2_0}\vec\psi_1\right]\\
	&\zero{-\frac{3\,\xi(\ell)\,\sigma^2_0}{(\sss)^2}\left[\vec\nabla\delta_1-\frac{\sigma^2_1}{\sigma^2_0}\,\vec\psi_1\right]^2 -\frac{\xi(\ell)\sigma^2_1}{\sigma^2_0(\sss)}}\Biggl\}\\
	+&\muBCD \Biggl\{ \,\frac{3\delta_1(\sigma^4_0+\sigma^2_1\sigma^2_{-1})}{\sigma^2_0(\sigma^4_0 - \sigma^2_1\sigma^2_{-1})^2}\,\vl \cdot \left(\frac{1}{\ell}\frac{\d \xi}{\d \ell}\,\vec\psi_1-\eta(\ell)\,\vec\nabla\delta_1\right) + \frac{6\delta_1}{(\sigma^4_0 - \sigma^2_1\sigma^2_{-1})^2} \,\vl\cdot\left[ \sigma^2_1\eta(\ell)\vec\psi_1- \sigma^2_{-1}\frac{1}{\ell}\frac{\d \xi(\ell)}{\d \ell} \vec\nabla\delta_1\right] \\
	&\zero{+\frac{3\xi(\ell)}{\sigma^2_0(\sss)^2}\,\left[\sigma^2_0(\sigma^2_1|\vec\psi_1|^2+\sigma^2_{-1}|\nabla\delta_1|^2) - (\sigma^4_0+\sigma^2_1\sigma^2_{-1})\,\vec\psi_1\cdot\vec\nabla\delta_1\right]+ \frac{\xi(\ell)}{\sss}}\Biggl\} \\
	\zero{+}&\zero{\frac{1}{2}\,\muBDD\Biggl\{\frac{6\,\delta_1}{(\sss)^2}\,\left[\sigma^2_{-1}\frac{1}{\ell}\frac{\d \xi(\ell)}{\d \ell} +\sigma^2_0 \eta(\ell)\right]\,\vl\cdot\left[\frac{\sigma^2_{-1}}{\sigma^2_0}\vec\nabla\delta_1-\vec\psi_1\right]}\\
	 &\zero{- \frac{3\,\xi(\ell)\,\sigma^2_0}{(\sss)^2}\left[\frac{\sigma^2_{-1}}{\sigma^2_0}\vec\nabla\delta_1-\vec\psi_1\right]^2 -\frac{\xi(\ell)\sigma^2_{-1}}{\sigma^2_0(\sss)}\Biggl\} }\,.
\end{aligned}
\end{equation*}
\end{footnotesize}
After performing the integral prescribed in \eqref{forcepeak}, we obtain the result \eqref{eq:force peak result}.

In \eqref{eq:force peak result} there are two cumulants containing $\delta(\vrp)$, $\muABC$ and $\muABD$. We report here the result of the integrals
\begin{eqnarray}
\int\d \ell\,\frac{\vec \ell}{\ell}\cdot\Big<\delta(\vrp)\delta(\vr)\vec \psi(\vr)\Big>&=& -\frac 12 \mathcal K_1 +\frac{10}{21}\sigma_0^2 \sigma_{-1}^2\,, \label{mu123}\\
\int\d \ell\,\frac{\vec \ell}{\ell}\cdot\Big<\delta(\vrp)\delta(\vr)\vec \nabla \delta(\vr)\Big>&=& \,\frac{17}{21}\sigma_0^4\,, \label{mu124}
\end{eqnarray}
where $\mathcal K_1$ is defined in \eqref{K1,K2}.

%
\nocite*{} 
\bibliographystyle{JHEP}
\bibliography{biblio_vb}

\end{document}